\documentclass[preprintnumbers,nofootinbib,noshowpacs,eqsecnum,prd]{revtex4}

\usepackage{graphicx,epsfig}
\usepackage{amsmath,amssymb,bbold}
\usepackage{url}

\graphicspath{ {Plots/} }

\makeatletter       

\renewcommand{\p@subsection}{}

\makeatother

\DeclareMathOperator{\sign}{sign}

\newcommand{\Emiss}{\ensuremath{E\!\!\!\!\slash}}

\newcommand{\qw}{\ensuremath{\tilde{q}}}

\newcommand{\gw}{\ensuremath{\tilde{g}}}

\newcommand{\mqw}{\ensuremath{m_{\tilde{q}}}}
\newcommand{\mgw}{\ensuremath{m_{\tilde{g}}}}

\newcommand{\G}{\ensuremath{{g}}}
\newcommand{\Gt}{\ensuremath{\tilde{g}}}

\begin{document}

\title{ \hfill{\footnotesize{DESY 08-113}} \\[-2mm]
        \hfill{\footnotesize{PITHA 08/19}} \\[6mm]
TESTING THE MAJORANA NATURE OF GLUINOS AND NEUTRALINOS
}

\author{S.~Y.~Choi$^1$, M.~Drees$^{2,3,4}$, A.~Freitas$^{5,6}$, and
        P.~M.~Zerwas$^{7,8}$\\[-3mm]
        \mbox{ } }
\address{ $^1$ Department of Physics and RIPC, Chonbuk National University,
        Jeonju 561-756, Korea \\
          $^2$ Physikalisches Inst.~der Univ.~Bonn, D-53115 Bonn, Germany\\
          $^3$ School of Physics, KIAS, Seoul 130-012, Korea \\
          $^4$ Bethe Center of Theoretical Physics, Univ. Bonn, D-53115 Bonn,
        Germany \\
          $^5$ Department of Physics \& Astronomy, University of Pittsburgh, PA
               15260, USA\\
          $^6$ HEP Division, Argonne National Laboratory,
        Argonne, IL 60439, USA \\
        $^7$ Deutsches Elektronen-Synchrotron DESY, D-22603 Hamburg, Germany \\
        $^8$ Inst.~Theor.~Physik E, RWTH Aachen U, D-52074 Aachen, Germany }
\date{\today}

\begin{abstract}
{\it Gluinos and neutralinos, supersymmetric partners of gluons and neutral
electroweak gauge and Higgs bosons, are Majorana particles in the Minimal
Supersymmetric Standard Model [MSSM]. Decays of such self-conjugate particles
generate charge symmetric ensembles of final states. Moreover,
production channels of supersymmetric particles at colliders are
characteristically affected by the Majorana nature of particles
exchanged in the production processes. The sensitivity to the Majorana
character of the particles can be quantified by comparing the predictions
with Dirac exchange mechanisms. A consistent framework for introducing
gluino and neutralino Dirac fields can be designed by extending the N=1
supersymmetry of the MSSM to N=2 in the gauge sector. We examine to which
extent like-sign dilepton production in the processes $qq \to \qw \qw$
and $e^- e^- \to {\tilde{e}}^- {\tilde{e}}^-$ is affected by the exchange
of either Majorana or Dirac gluinos and neutralinos, respectively, at
the Large Hadron Collider (LHC) and in the prospective $e^- e^-$ mode of a
lepton linear collider.}
\end{abstract}

\maketitle

%
\section{Introduction}

\noindent
In the Minimal Supersymmetric Standard Model [MSSM] gauge super-multiplets are
built up by two components, bosonic gauge fields and fermionic gaugino fields,
Refs.~\cite{Wess,Nilles,Drees}. Since neutral vector fields are
self-conjugate, the corresponding supersymmetric partners are Majorana fields.
Condensing the gluon fields in the color-octet matrix $\G$ and the gluinos in
the color-octet matrix $\Gt$, the (color) charge conjugate fields ${\G}^c$ and
${\Gt}^c$ are related to the original fields by
\begin{eqnarray}
  {\G}^c  &=& - \, {\G}^T         \nonumber \\
  {\Gt}^c &=& - \, {\Gt}^T                  \,.
\end{eqnarray}
For the electroweak gauge and Higgs bosons and the neutralinos, mixtures of
fermionic gauginos and higgsinos, analogous relations hold.

The gluino and neutralino Majorana particles carry masses which are rooted in
the Higgs and the (soft) supersymmetry breaking sector. Massive Majorana fields
can be distinguished experimentally from Dirac fields in gauge theories quite
generally.  [For massless fields the distinction is more subtle, depending on
the form of the interactions in the theory.] In this report we will study the
characteristic differences between Majorana and Dirac fields and work out the
experimental implications. The analyses will be performed in a hybrid scheme,
Ref.~\cite{Benakli}, in which the minimal N=1 supersymmetric standard
model is extended by gauge elements of N=2 supersymmetry \cite{Fayet}.

Majorana fields in N=1 supersymmetric theories are characterized by two
self-conjugate L- and R-components in parallel to the two vector field
components.  These fermionic components can be paired with two additional
fermionic fields in N=2 supersymmetric theories, in which a vector
super-multiplet is combined with an additional chiral super-multiplet to a
vector hyper-multiplet. If the Majorana masses are identical and the fields
are mixed maximally, the four fermionic degrees of freedom can join to a Dirac
field and its charge-conjugate companion \cite{Majorana_to_Dirac}. In this
limit the theory includes vector fields, Dirac gaugino fields and scalars, all
states belonging to the adjoint representation of the gauge group.

The four Higgs superfields belong to a chiral and an anti-chiral multiplet.
By contrast, the matter superfields {\it sui generis} are restricted in the
N=1/N=2 hybrid scheme to the standard N=1 chiral component in accordance with
the experimental fact that matter fermions are chiral.

In this setup the N=2 gauge interactions are an extension of the familiar N=1
gauge interactions. The additional component of the N=2 interaction between
the new gaugino field and the Higgs fields can be reinterpreted as component
of a superpotential affecting the neutralino and chargino masses after the
electroweak symmetry is broken. [In addition the Higgs self-interactions are
modified, not affecting the present analysis though.]

Soft supersymmetry breaking gives rise, at the phenomenological level, to
three gaugino mass parameters: two Majorana masses $M_a$ and $M_b$, and a
mixing term $M_{ab}$. The first Majorana mass may be associated with the N=1
gaugino mass term, and the second with the new gaugino field.  Diagonalizing
the $\{ab\}$ mass matrix generates two Majorana masses $m_{1,2}$ and a mixing
angle $\theta$, which relates the mass eigenstates to the original current
states. Depending on the supersymmetry breaking parameters, $\theta$ can
assume any value between 0 and $\pi/2$. It is easy to design a path in
mass-parameter space such that the Dirac limit can be approached smoothly.
Tuning the diagonal mass parameters $M_{a,b}$ to zero, only the off-diagonal
mixing term $M_{ab}$ survives and, as a result, the mass eigenvalues $m_{1,2}$
become identical, {\it modulo} sign, and the mixing of the states maximal,
$\theta = \pi/4$. In the maximal mixing limit the two Majorana states combine
into one Dirac fermion (and its antifermion partner).  Maximal mixing of
Majorana particles guarantees the vanishing of transition amplitudes generally
associated with the exchange of Dirac particles.

This procedure is well suited for the strong interaction sector. The
electroweak sector is less transparent due to the complicated mixing effects
beyond the soft supersymmetry breaking terms after electroweak symmetry
breaking. In the limit in which the supersymmetry breaking scale is
significantly larger than the electroweak scale, the Dirac limit is approached
approximately. Though sounding strange at first glance, it is clear in the
light of the previous comments that a quantitative definition can be
formulated for the concept of a near-Dirac field or particle.

Adopting this extension of the MSSM to a N=1/N=2 hybrid model, observables can
be designed for experimental analyses at the LHC \cite{LHC}, which allow us to
follow a smooth transition from a Majorana theory of gluinos (and neutralinos)
to a Dirac theory. The standard examples are the equal-chirality transition
amplitudes
\begin{equation}
q_L q_L \to {\tilde{q}}_L {\tilde{q}}_L  \;\;\; {\rm and} \;\;\;
q_R q_R \to {\tilde{q}}_R {\tilde{q}}_R                           \,.
\end{equation}
These amplitudes are non-zero for Majorana gluino exchange but they vanish for
Dirac gluino exchange in the N=1/N=2 hybrid theory. [The same arguments can be
applied to $e^-_L e^-_L \to {\tilde{e}}^-_L {\tilde{e}}^-_L$, and L$
\Rightarrow$ R, for electroweak gauginos.] However, amplitudes for the
transition from 2-fermion to 0-fermion states do not vanish in general. In the
present context the mixed-chirality amplitude $q_L q_R \to {\tilde{q}}_L
{\tilde{q}}_R$ is non-zero for Dirac exchange and, in fact, equal to the
amplitude for Majorana exchange [analogously for $e^-_L e^-_R$ scattering].

Using left/right-handedly polarized beams in the $e^-e^-$ collision mode of a
linear collider \cite{LC}, the rules outlined above can easily be applied for
studying the Majorana/Dirac nature of neutralinos experimentally. In addition,
it has been demonstrated earlier when discussing potential measurements of the
$q\tilde{q}\tilde{g}$ Yukawa coupling, Refs.~\cite{yuk,yuk2}, that the
analysis of like-sign dilepton final states in $pp$ collisions at the LHC
signals ${\tilde{q}}_L {\tilde{q}}_L$ final states in supersymmetric theories.
Adjusting the ${\tilde{q}}_L$ decays to the Dirac limit, the analyses of
Refs.~\cite{yuk,yuk2} can be transferred, {\it mutatis mutandis}, easily. The
potential of like-sign dilepton signatures for discriminating Majorana from
Dirac structures of supersymmetric theories has also been noted in
Ref.~\cite{no}.

These processes are complementary to tests of the Majorana nature of gluinos
in gluino decays, notably to top plus stop final states, which have been
discussed widely in the literature \cite{gluino_Majorana}.  Moreover,
two-gluino final states decaying to bottom + sbottom quarks have served as an
important channel for searching for supersymmetry at the Tevatron \cite{Tev}.
Likewise, analyses of like-sign chargino production \cite{alwall} at the LHC and
neutralino decays \cite{neutralino_decay_Majorana} have been studied
extensively in the past for testing the Majorana character of neutralinos.

The report is organized as follows. In the next Section 2 we define the
essential elements of the N=1/N=2 hybrid model and establish the
phenomenological base. In Sections 3 and 4 we discuss subsequently the strong
interaction gluino sector and the electroweak sector, including the concept of
a near-Dirac field, first in the limit in which the electroweak breaking scale
can be neglected compared to the supersymmetry parameters, and second the
systematic approximation to this limit. In Section 5 like-sign dileptons
will be analyzed as a signal for the Majorana to Dirac transition at the LHC,
before Section 6 concludes this study.

%
\section{Theoretical Basis: N=1/N=2 Hybrid Model}

\noindent
In the MSSM based on N=1 supersymmetry, bosonic gauge fields are one-to-one
paired with fermionic spin 1/2 gaugino fields and Higgs bosons with higgsinos.
Fermionic lepton and quark matter fields are paired with bosonic spin-0
sleptons and squarks. The neutral gauginos in this ensemble are self-conjugate
Majorana fields with two chirality components corresponding to the two
helicity states of the gauge fields. Decay channels of these particles and
their exchange in production processes generate characteristic signatures of
their Majorana nature.  The uniqueness of these characteristics can be proven
by comparing the signatures with predictions derived from Dirac theories.

It turns out that N=2 supersymmetry offers a theoretically solid platform for
a consistent comparison between Majorana and Dirac theories
\cite{Benakli,supersoft}.
The gauge super-multiplets are expanded to hyper-multiplets which incorporate
new chiral superfields composed of a gaugino and a scalar field. In the
following these new gaugino fields will be labeled by an apostrophe. For the
standard SU(3)$\times$SU(2)$\times$U(1) gauge group the N=2 fields and their
components are summarized in Table~\ref{tab:hyper}. As argued before, the
superposition
\begin{table}
\begin{tabular}{|c||c|c|c|}
\hline
Group  &  Spin 1     & Spin 1/2                   & Spin 0         \\
\hline\hline
SU(3)  & $g$         & $\tilde{g}; {\tilde{g}}'$  & $\sigma_g$     \\
SU(2)  & $W^\pm,W^0$ & ${\tilde{W}}^\pm,{\tilde{W}}^0;
            {\tilde{W}}^{\prime\pm},{\tilde{W}}^{\prime 0}$ &
                                    ${\sigma}_W^\pm,{\sigma}_W^0$    \\
U(1)   & $B$         & $\tilde{B};{\tilde{B}'}$   & $\sigma_B$     \\
\hline
\end{tabular}
\caption{The N=2 gauge hyper-multiplets.}
\label{tab:hyper}
\end{table}
of two Majorana fields carrying equal masses and being mixed maximally can be
reinterpreted as a Dirac field. By tuning the masses of the N=1 gauginos and
the new gauginos a path for a continuous transition from a Majorana to a Dirac
theory can be designed.

The two disjoint N=1 super-fields of the MSSM Higgs sector ${\hat{H}}_d$ and
${\hat{H}}_u$ can be united in an N=2 hyper-multiplet, composed of
${\hat{H}}_d$ as a chiral field and ${\hat{H}}^\dagger_u$ as its anti-chiral
companion, cf. Ref.$\,$\cite{QA}.

In a similar way the chiral matter superfields of (s)leptons and (s)quarks,
generically called $\hat{Q}$, are extended by new anti-chiral matter fields
${\hat{Q}}'$ to hyper-fields. None of the mirror fields ${\hat{Q}}'$ that
include new leptons and quarks has been observed so far.  Given the success of
the chiral standard theory, either the mirror particles are very heavy, or
this component is assumed absent {\it a priori}. The second scenario may be
realized in N=2 theories including extra space dimensions in which N=1 matter
super-fields are restricted only to 4-dimensional branes \cite{Chacko:2004mi}.

Alternative supersymmetric scenarios with Dirac gauginos are based on D-term
supersymmetry breaking models \cite{dterm} or exact continuous R-symmetries
\cite{rsymm}. On the phenomenological level these models lead to identical
formulations of the Dirac gauginos
but it is less straightforward to define a continuous Majorana-Dirac
transition.

In the following we will adopt the N=1/N=2 hybrid scenario as the base for
phenomenological studies of smooth transitions from Majorana to Dirac fields.
The model appears minimal in view of the basic field degrees of freedom and
their interactions. For the present purpose there is little difference between
the N=2 form of the Higgs sector or two disjoint N=1 Higgs sectors treated in
parallel to the matter fields. [An increased mass range of the lightest Higgs
boson and additional self-couplings however render the extended option
attractive in itself.]

Concentrating on the gaugino sector in regard of the Majorana to Dirac
transition, the Lagrangian derived from the general N=2 action can be
restricted to a few relevant terms:

\subsection{Hyper-QCD Sector}

\noindent
Standard gluino $\tilde{g}$ and new gluino ${\tilde{g}}'$ fields are coupled
minimally to the gluon field $g$,
\begin{equation}
    {\mathcal{L}}_{\rm QCD}^{g \gw \gw} = g_s {\rm Tr}\,
    \bigl ( \overline{\Gt} \gamma^{\mu} [{\G}_{\mu}, {\Gt}] +
          \overline{\Gt'} \gamma^{\mu} [{\G}_{\mu}, {\Gt}'] \bigr ) \,,
  \label{gluinogauge}
\end{equation}
with the fields condensed to color-octet matrices ${\G}_\mu =
\frac{1}{\sqrt{2}} \lambda^a g^a_\mu$ etc., $g_s$ denoting the QCD coupling,
and two 4-component Majorana spinor fields $\Gt$ and $\Gt'$ satisfying
$(\Gt)^c = -\Gt^T$ and $(\Gt')^c= -\Gt'^T$. The Lagrangian generates the usual
$\tilde{g} \tilde{g} g$ and $\tilde{g}' \tilde{g}' g$ vertices for gluinos
coupled to gluons. Matter fields only interact with the standard gluino,
\begin{equation} \label{qcd-yuk}
   \mathcal{L}_{\rm QCD}^{q\tilde{q}\tilde{g}}
   = -  g_s \left[\,\overline{q_L}  \gw \, \tilde{q}_L
                 - \overline{q_R}  \gw \, \tilde{q}_R
                 + {\rm h.c.}\right] \,,
\end{equation}
while N=2 supersymmetry requires $\tilde{g}'$ to only couple to the
hyper-multiplet partners of the N=1 quarks/squarks which, in the hybrid
theory, are assumed to be projected out.\footnote{One could contemplate a
  non-supersymmetric theory with Dirac gluinos where $\tilde g' \bar q \tilde
  q$ couplings exist. This would tend to increase the differences between
  Majorana and Dirac gluinos, e.g. leading to different total cross sections
  for associate $g q \rightarrow \tilde q \tilde g$ production. The N=1/N=2
  hybrid analyzed by us is better motivated; considering it as alternative of
  the usual MSSM is also conservative in the sense that it minimizes the
  differences.}

Soft supersymmetry breaking generates masses for the gluino fields $\tilde{g}$
and $\tilde{g}'$. Diagonal terms in the fields $\tilde{g}$ and ${\tilde{g}}'$
generate the individual Majorana mass parameters $M_3$ and $M_3'$ while an
off-diagonal term coupling $\tilde{g}$ with ${\tilde{g}}'$ will be crucial for
the transition of the two Majorana fields to a joined Dirac field:
\begin{equation}
    {\mathcal{L}}_{\rm QCD}^m
    =   - \frac{1}{2}
        \left[ M_3'\, {\rm Tr}(\overline{\Gt'}  {\Gt}')
        + M_3\, {\rm Tr}(\overline{\Gt}  \Gt)
        + M_3^D\, {\rm Tr}(\overline{\Gt'} \Gt +
                         \overline{\Gt}  \Gt' ) \right]
           \,.
\end{equation}
[For the purpose of our analysis, all mass parameters are assumed real
throughout the paper.] As worked out in detail in the next section,
diagonalizing the $\tilde{g}',\tilde{g}$ mass matrix [in the left-chirality
basis, i.e.  $\tilde{g}_L = \frac{1}{2}(1 - \gamma_5) \tilde{g}$ etc.]
\begin{equation}
   {\mathcal{M}}_g = \begin{pmatrix}
                     M'_3   &   M^D_3  \\
                     M^D_3  &   M_3
                     \end{pmatrix}
\label{eq:gluino_mass_matrix}
\end{equation}
gives rise to two Majorana mass eigenstates, $\tilde{g}_1$ and $\tilde{g}_2$
with masses $m_1$ and $m_2$. For large new gluino masses, $M_3' \to \pm
\infty$, the standard MSSM gluino sector is recovered.  On the other side, in
the limit in which the Majorana mass parameters $M_3$ and $M_3'$ vanish but
the off-diagonal element $M^D_3$ is non-zero, the mixing between the states is
maximal and the two Majorana states, carrying identical masses, can be paired
to a Dirac state.  Thus varying $M_3'$ from infinity to zero while trailing
$M_3$ from a TeV-scale value to zero, a continuous path can be constructed for
the transition from the MSSM gluino Majorana theory to a Dirac theory.

Table I shows that the hybrid theory also contains a complex scalar octet
$\sigma_g$. Its coupling to gluons is determined by SU(3) gauge invariance.
In addition, N=2 supersymmetry stipulates \cite{Fayet} the existence of a
$\sigma_g \Gt \Gt'$ coupling, while the couplings of $\sigma_g$ to quarks also
involve their hyper-multiplet partners. The hybrid theory predicts pair
production of $\sigma_g$ scalars. However, this is not directly related to the
Dirac or Majorana nature of the gluinos, which is the central issue of our
analysis. The detailed phenomenology of the new scalars will be described
in a sequel to this report.

\subsection{Electroweak Sector}

\noindent
The electroweak neutralino/chargino sector is considerably more complicated
than the QCD sector due to the mixing of gauginos and higgsinos induced by
electroweak symmetry breaking. The complexity increases only slightly in the
extension from N=1 to N=2 supersymmetry. While the expansion of the
$\tilde{W},\tilde{B}$ isospin and hypercharge sector by the
${\tilde{W}}',{\tilde{B}}'$ fields runs strictly parallel to the gluino
sector, the embedding of the Higgs fields into a chiral and anti-chiral N=2
hyper-multiplet generates new gauge interactions which couple the Higgs
super-fields with the new chiral superfields of the N=2 vector multiplets:
\begin{equation}
    W_{\rm higgs}^{\rm gauge'}
   = \sqrt{2}\, g\, \hat{H}_u\cdot (I^a \hat{H}_d)\, \hat{W}^{\prime a}
    +\sqrt{2}\, g'\, \hat{H}_u \cdot (Y \hat{H}_d)\, \hat{B}' \,,
\label{newgauge}
\end{equation}
where $I^a = \tau^a/2$ ($a=1,2,3$) and $Y$ are the weak isospin and
hypercharge generators, respectively, $g$ and $g'$ are the SU(2) and
U(1)$_{\rm Y}$ gauge couplings, and the central dot denotes an SU(2)-invariant
contraction. The N=2 supersymmetry allows for a bilinear $\mu$
Higgs/higgsino coupling,
\begin{equation}
   W_{\rm higgs}^{\rm bilin} = \mu \hat{H}_u \cdot \hat{H}_d
\end{equation}
[in the standard notation with the SU(2)-invariant contraction
$\hat{H}_u\cdot \hat{H}_d = \hat{H}^+_u \hat{H}^-_d - \hat{H}^0_u \hat{H}^0_d$,
etc].

The additional gauge-strength Yukawa interactions,
\begin{equation}
   {\cal L}^{\rm gauge'}_{\rm higgs}
   =  - \frac{g}{\sqrt{2}}
    \left[ H_u\cdot (\tau^a \tilde{H}_d)\, \tilde{W}^{\prime a}
         + H_d\cdot (\tau^a \tilde{H}_u)\, \tilde{W}^{\prime a} \right]
    - \frac{g'}{\sqrt{2}}
    \left[ H_d\cdot \tilde{H}_u\, \tilde{B}^\prime
         - H_u\cdot \tilde{H}_d\, \tilde{B}^\prime \right] \,,
\end{equation}
generated from the superpotential Eq.~\eqref{newgauge}, lead, after
electroweak symmetry breaking,
\begin{align} \label{newgy}
&\mbox{Neutralinos:} & {\mathcal{L}}_{\rm higgs}^{\rm \chi'^0}
&= - m_Z
  \left[ s_W ( s_\beta \overline{\tilde{B}'_R}  \tilde{H}^0_{dL}
             + c_\beta \overline{\tilde{H}^0_{uR}} \tilde{B}_L' )
       - c_W ( c_\beta \overline{\tilde{H}^0_{uR}} \tilde{W}'^0_L
             + s_\beta \overline{\tilde{W}'^0_R} \tilde{H}^0_{dL} )
+ \mbox{h.c.}\right],\\
&\mbox{Charginos:} &  {\mathcal{L}}_{\rm higgs}^{\rm \chi'^\pm}
&= - \sqrt{2} m_W c_\beta \overline{\tilde{H}^-_{uR}} \tilde{W}'^-_L
   + \sqrt{2} m_W s_\beta \overline{\tilde{W}'^-_R} \tilde{H}^-_{dL} 
+ \mbox{h.c.},
\end{align}
to off-diagonal mass terms and mixings between the standard higgsinos and the
new winos ${\tilde{W}}'$ and bino ${\tilde{B}}'$. Choosing the left-chirality
bases $ \{ {\tilde{B}}'^0, {\tilde{B}}^0, {\tilde{W}}'^0, {\tilde{W}}^0,
{\tilde{H}}^0_d, {\tilde{H}}^0_u \}$ and $\{ {\tilde{W}}'^\mp,
{\tilde{W}}^{\mp}, {\tilde{H}}^{\mp}_{d,u} \}$, the neutralino and chargino
mass matrices can be cast into the form
\begin{align}
\label{eq:neu_mass_matrix}
{\mathcal{M}}_n  &= \left(\begin{array}{c c c c c c}
 M'_1  & M^D_1 & 0     & 0     &  m_Z s_W s_\beta &   m_Z s_W c_\beta  \\
 M^D_1 & M_1   & 0     & 0     & -m_Z s_W c_\beta &   m_Z s_W s_\beta  \\
 0     & 0     & M'_2  & M^D_2 & -m_Z c_W s_\beta & - m_Z c_W c_\beta  \\
 0     & 0     & M^D_2 & M_2   &  m_Z c_W c_\beta & - m_Z c_W s_\beta  \\
 m_Z s_W s_\beta & -m_Z s_W c_\beta & -m_Z c_W s_\beta &  m_Z c_W c_\beta & 0
 & -\mu \\
 m_Z s_W c_\beta &  m_Z s_W s_\beta & -m_Z c_W c_\beta & -m_Z c_W s_\beta 
      & -\mu & 0    \\
\end{array}\right),  \\[3ex]
{\mathcal{M}}_c  &= \left(\begin{array}{c c c}
 M'_2   & M^D_2 & -\sqrt{2} m_W \sin \beta \\
 M^D_2 & M_2   & \sqrt{2} m_W \cos \beta \\
 \sqrt{2} m_W \cos \beta & \sqrt{2} m_W \sin \beta & \mu \\
\end{array}\right) ,
\end{align}
with the usual abbreviations $s_W = \sin\theta_W$, $s_\beta = \sin\beta$, 
etc. for the electroweak mixing angle $\theta_W$ and the SUSY
Higgs-Goldstone mixing angle $\beta$. [Evidently, the new N=2 Higgs-gauge
interactions (\ref{newgauge}) have little impact on the overall structure of
the mass matrices.  If the Higgs sector is reduced to the standard twin of
N=1 Higgs fields, the terms corresponding to Eq.\eqref{newgy} are simply
reduced to zero.]

The gaugino-gauge interactions are extended analogously to the gluino-gluon
sector in Eq.~\eqref{gluinogauge}. A new set of interactions between Higgs,
higgsino, gauge and gaugino fields is generated  by the N=2 gauge interactions
in the Higgs sector, cf. Eq.\eqref{newgauge}.

The superpotential involving matter and Higgs superfields of the hybrid model
will be taken over from N=1 supersymmetry, analogously the corresponding soft
supersymmetry breaking interactions.

%
\section{The Gluino Sector in Super- and Hyper-QCD}
\label{sc:hyperqcd}

\noindent
In the previous section we have derived the mass matrix in the gluino sector
of the two Majorana fields $\gw$ and ${\gw}'$ in N=2 supersymmetry. In the
present section we will determine the mass eigenvalues and the corresponding
gluino fields. Two limiting cases of the general softly broken N=2 theory
are of particular interest. If one of the Majorana mass parameters in the
gluino mass matrix is driven to infinity, we will recover the standard N=1
supersymmetry. On the other hand, if both diagonal mass parameters are chosen
zero, the two Majorana fields can be united to a Dirac field. This transition
restricts considerably the non-zero scattering amplitudes generated by gluino
exchanges. Thus by tuning the mass parameters, a common platform for Majorana
and Dirac theories can be built, allowing for continuous transitions between
the two types of fields and a proper definition of a ``near-Dirac'' field.

\subsection{Diagonalization of the 2$\times$2 Hyper-Gluino Mass Matrix}

\noindent
For real values of $M_3, M'_3$ and $M^D_3$ the gluino mass matrix in
Eq.~\eqref{eq:gluino_mass_matrix} can be diagonalized by means of the unitary
transformation matrix $\mathcal{U}$,
\begin{equation}
 \mathcal{U}^T{\cal M}_g \, \mathcal{U}
 = {\rm diag}(m_{\tilde{g}_1}, m_{\tilde{g}_2}) \quad {\rm with} \quad
\mathcal{U} = \left(\begin{array}{cc}
      \cos\theta_3  &  \epsilon_3 \sin\theta_3 \\
      -\epsilon_3 \sin\theta_3 & \cos\theta_3 \end{array}
      \right)
      \left(\begin{array}{cc}
      \eta_1 & 0  \\
      0 & \eta_2 \end{array}
      \right)\,,
\label{eq:gluino_mass_matrix_diagonalization}
\end{equation}
where the rotation angle varies between $0\leq \theta_3\leq \pi/2$ and
$\eta_{1,2}$ denote the two Majorana-type phases.  The mass eigenvalues read:
\begin{equation}
m_{\tilde{g}_{1,2}}
  = \frac{1}{2} \left||M'_3+M_3| \mp \Delta_3\right|
    \quad {\rm with} \quad \Delta_3 = \sqrt{(M'_3-M_3)^2
+ 4 (M^D_3)^2}\,,
\end{equation}
with the ordering $m_{\tilde{g}_1}\leq m_{\tilde{g}_2}$ by definition. The
mixing angle $\theta_3$, the sign parameter $\epsilon_3$ and the two
Majorana-type phases $\eta_{1,2}$ defining the diagonalization matrix
$\mathcal{U}$ are given by
\begin{eqnarray}
&& \cos\theta_3/\sin\theta_3
   = \sqrt{\frac{1\pm\epsilon'_3 |M_3-M'_3|/\Delta_3}{2}}      \,,\nonumber\\
&& \epsilon_3 = \sign[M_3^D (M_3'-M_3)] \;\;\; {\rm and} \;\;\;
   \epsilon'_3 =\sign[M^2_3-M'^2_3] \,,
\end{eqnarray}
and
\begin{eqnarray}
&& \eta_1 =  1/i \quad  \mbox{for}\quad
   \sign [{\rm det}({\mathcal{M}}_g) \cdot {\rm Tr}({\mathcal{M}}_g)]= +/- \,,
\nonumber\\
&& \eta_2 =  1/i \quad  \mbox{for}\quad
   \sign [ {\rm Tr}({\mathcal{M}}_g)] = +/-
\end{eqnarray}
[the overall signs of $\eta_{1,2}$ are indeterminate].
The form of the diagonalization matrix $\mathcal{U}$ in
Eq.~\eqref{eq:gluino_mass_matrix_diagonalization} guarantees the positivity
of the mass eigenvalues $m_{{\gw}_{1,2}}$ of the fields
\begin{equation}
\begin{pmatrix}
    \gw_{1R}  \\
    \gw_{2R}
\end{pmatrix}
   =
\mathcal{U}^T
\begin{pmatrix}
   \tilde{g}'_R  \\
   \tilde{g}_R
\end{pmatrix}
\;\;\; {\rm and} \;\;\;
\begin{pmatrix}
    \gw_{1L}  \\
    \gw_{2L}
\end{pmatrix}
   =
\mathcal{U}^\dagger
\begin{pmatrix}
   \tilde{g}'_L  \\
   \tilde{g}_L
\end{pmatrix}    \,.
\end{equation}
The rotation by means of the orthogonal sub-matrix of $\mathcal{U}$, combined
with the diagonal phase matrix, preserves the Majorana character of the fields
${\gw}_{1,2}$. Left- and right-chiral fields are related by charge
conjugation: $(\tilde{g}_L)^c = -\tilde{g}_R^T$ and $(\tilde{g}'_L)^c =
-\tilde{g}'^T_R$.

\subsection{Chiral Transition Amplitudes in the Hybrid Model}

\begin{figure}
\psfig{figure=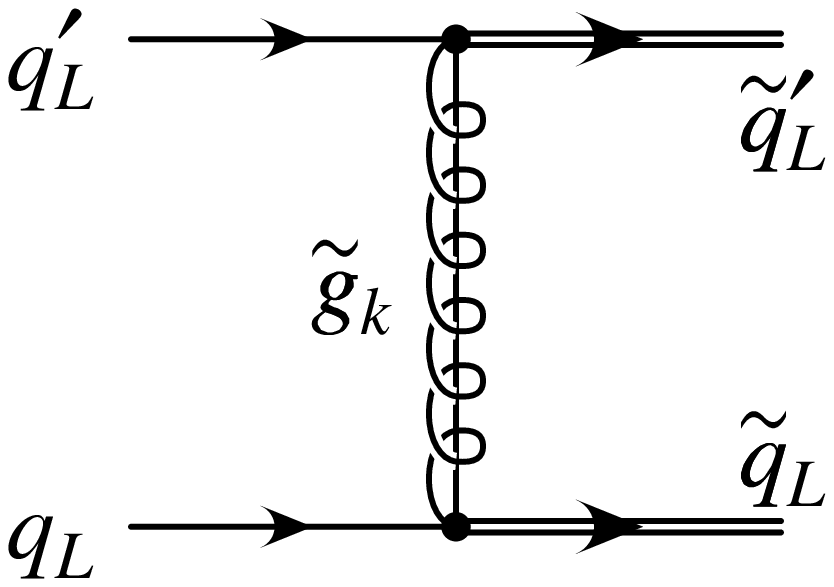, width=4cm}
\psfig{figure=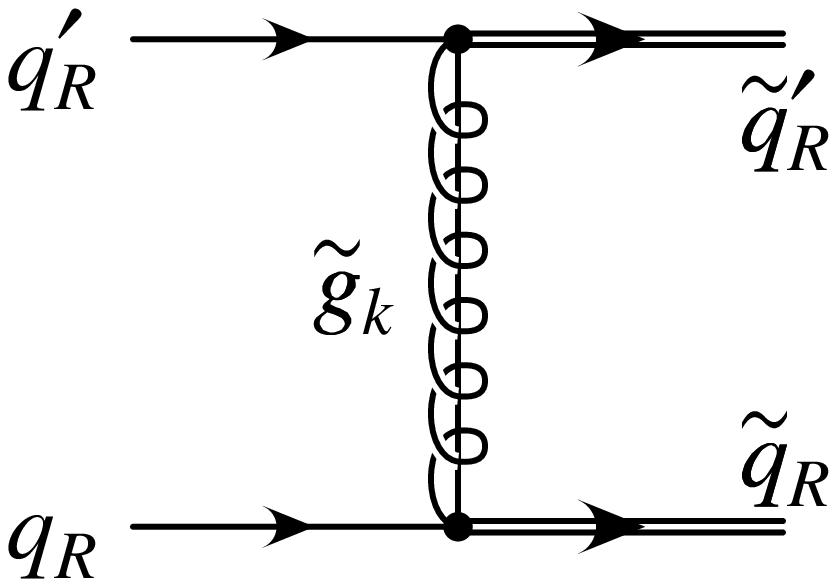, width=4cm}
\psfig{figure=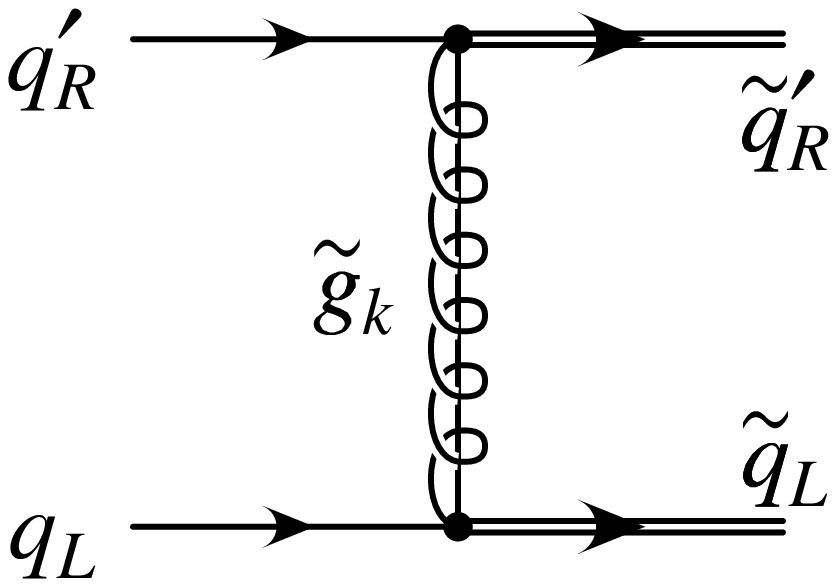, width=4cm}\\[-2ex]
\makebox[4cm]{(a)}
\makebox[4cm]{(b)}
\makebox[4cm]{(c)}
\caption{Feynman diagrams for squark production of different flavors at hadron
colliders. [The index $k$ counts the two gluinos in the N=2 hybrid model, to be
ignored for the N=1 MSSM.]}
\label{fig:dia1}
\end{figure}

\noindent
The most transparent example for studying the Majorana/Dirac nature of gluinos
is the transition between pairs of quarks to pairs of squarks with different
flavor:
\begin{eqnarray}
  q_L q'_L &\to& \qw_L \qw_L'\;,\;\; q_R q'_R \to \qw_R \qw_R' \,,  
\\ \nonumber
  q_L q'_R &\to& \qw_L \qw_R'  \,,
\end{eqnarray}
The corresponding Feynman diagrams are depicted in Fig.~\ref{fig:dia1}.
The kernels of the transition matrix elements, involving the $t$-channel
exchange of the two Majorana gluinos ${\gw}_k$, can be cast into the form:
\begin{eqnarray}
  A[q_L q_L' \to \qw_L \qw_L'] &=&
    - \frac{g^2_s}{2} \sum_{k=1}^2
      \left[\mathcal{U}_{2k} \mathcal{U}_{2k}\right]
      \frac{\langle m_{\tilde{g}_k}\rangle}{t-m^2_{\tilde{g}_k}}       \,,
    \nonumber\\
  A[q_R q_R' \to \qw_R \qw_R'] &=&
   - \frac{g^2_s}{2} \sum_{k=1}^2
     \left[ \mathcal{U}^\ast_{2k} \mathcal{U}^\ast_{2k} \right]
     \frac{\langle m_{\tilde{g}_k}\rangle}{t-m^2_{\tilde{g}_k}}   \,, 
 \nonumber \\
   A[q_L q_R' \to \qw_L \qw_R'] &=&
   + \frac{g^2_s}{2} \sum_{k=1}^2 \left[ \mathcal{U}_{2k}
     \mathcal{U}^\ast_{2k} \right]
     \frac{\langle\not\!{q}\rangle}{t-m^2_{\tilde{g}_k}}                \,,
\label{eq:transition_amplitude}
\end{eqnarray}
to be sandwiched between the quark spinors $\bar{v}'_{L,R}$ and $u_{L,R}$; $t
= q^2$ denotes the square of the momentum $q$ flowing through the gluino line.
The form of these transition amplitudes can easily be traced back to the rules
introduced in the previous section. Currents of equal-sign chirality, LL and
RR, are coupled by the mass term of the gluino propagator, while currents of
opposite-sign chirality, LR, are coupled by the kinetic term $q$.

From the transition amplitudes (\ref{eq:transition_amplitude}) the cross
sections can easily be derived as
\begin{eqnarray}
\label{eq:xsec_diff_fl}
  \sigma[qq' \to \qw_L \qw_L'] &=& \sigma[qq' \to \qw_R \qw_R']
                                        \\[1.5ex]  \nonumber
                                    &=&
\frac{2 \pi \alpha_s^2 }{9 \, s} \biggl [
\frac{c_3^4 \, s \, \beta \, m_{\tilde{g}_1}^2}{s \, m_{\tilde{g}_1}^2 +
    (m_{\tilde{g}_1}^2 - m_{\tilde{q}}^2)^2}
+ \frac{s_3^4 \, s \, \beta \, m_{\tilde{g}_2}^2}{s \, m_{\tilde{g}_2}^2 +
    (m_{\tilde{g}_2}^2 - m_{\tilde{q}}^2)^2}
+ \frac{2 c_3^2 s_3^2 \,m_{\tilde{g}_1} \, m_{\tilde{g}_2}}{m_{\tilde{g}_1}^2 -
    m_{\tilde{g}_2}^2} \; (L_1-L_2)
\biggr ]
       \\[3ex] \nonumber
  \sigma[qq' \to \qw_L \qw_R'] &=& \frac{2 \pi \alpha_s^2}{9 \, s}
   \biggl [ c_3^4 \Bigl ( \Bigl (1 +
    \frac{2}{s}(m_{\tilde{g}_1}^2  - m_{\tilde{q}}^2)\Bigr)
  L_1 - 2 \beta \Bigr)
+ s_3^4 \Bigl (\Bigl (1 +
    \frac{2}{s}(m_{\tilde{g}_2}^2  - m_{\tilde{q}}^2)\Bigr)
  L_2 - 2\beta \Bigr )
       \\ \nonumber
&& \hspace{3em} + 2 c_3^2 s_3^2  \biggl (
  \frac{(s \, m_{\tilde{g}_1}^2 + (m_{\tilde{g}_1}^2 - m_{\tilde{q}}^2)^2) L_1
    - (s \, m_{\tilde{g}_2}^2 + (m_{\tilde{g}_2}^2 - m_{\tilde{q}}^2)^2) L_2}
    {m_{\tilde{g}_1}^2 - m_{\tilde{g}_2}^2} -\beta \biggr)\biggr ]
\end{eqnarray}
where
\begin{equation}
   L_k = \log\frac{\! (1+\beta)+2 (m_{\tilde{g}_k}^2 -  m_{\tilde{q}}^2)/s}
         {\; (1-\beta)+2( m_{\tilde{g}_k}^2 - 2 m_{\tilde{q}}^2)/s}
\end{equation}
and $s_3=\sin \theta_3$, $c_3=\cos \theta_3$ and $\beta = (1-4
m_{\tilde{q}}^2/s)^{1/2}$; it has been assumed that all squarks have the same
mass $m_{\tilde{q}}$. In the next subsection the characteristics of the
transition amplitudes will be analyzed in detail.

\subsection{Majorana to Dirac Path in the Hybrid Model}

\noindent
The N=2 gluino mass matrix ${\mathcal{M}}_g$ is defined by three parameters,
two on-diagonal Majorana mass parameters and the off-diagonal mass parameter
which couples the two N=1 sectors of the gluino hyper-multiplet. In the
physical basis they manifest themselves as two Majorana mass eigenvalues
$m_{{\gw}_{1,2}}$ and the rotation angle $\theta_3$ between the current and
mass eigenstates.

If the new gluino mass parameter $M_3'$ is chosen infinitely large, the
hyper-system is reduced effectively to the original N=1 gluon-gluino
super-multiplet with the gluino mass determined by $M_3$,
\begin{eqnarray}
  m_{{\gw}_1} &\simeq& |M_3 - (M^D_3)^2/M_3'|  \to |M_3|   \nonumber \\
  m_{{\gw}_2} &\simeq& |M_3'| \hspace{6.2em}   \to \infty   \,,
\end{eqnarray}
in analogy to the seesaw formula.

\begin{figure}
\epsfig{figure=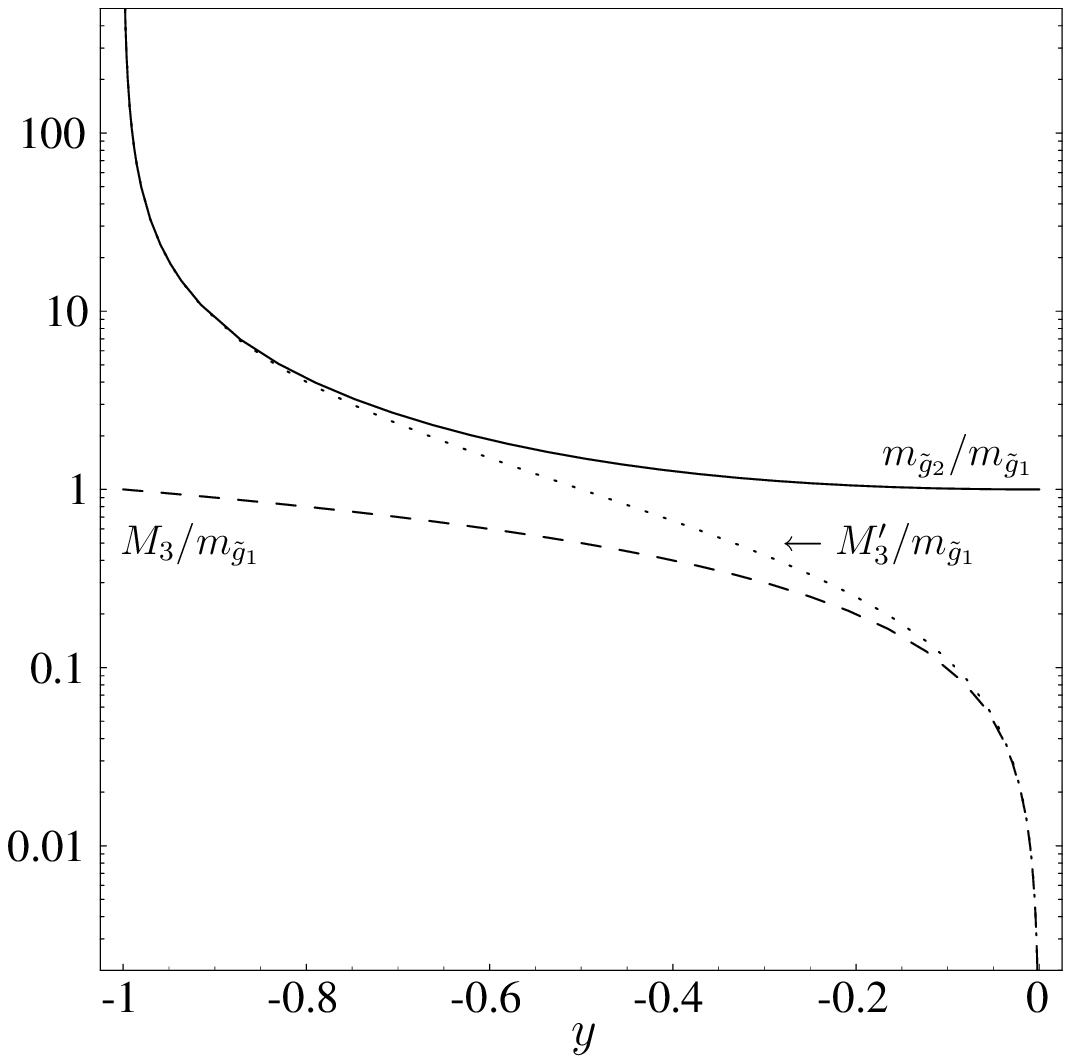, height=6.4cm}
\hspace{1cm}
\epsfig{figure=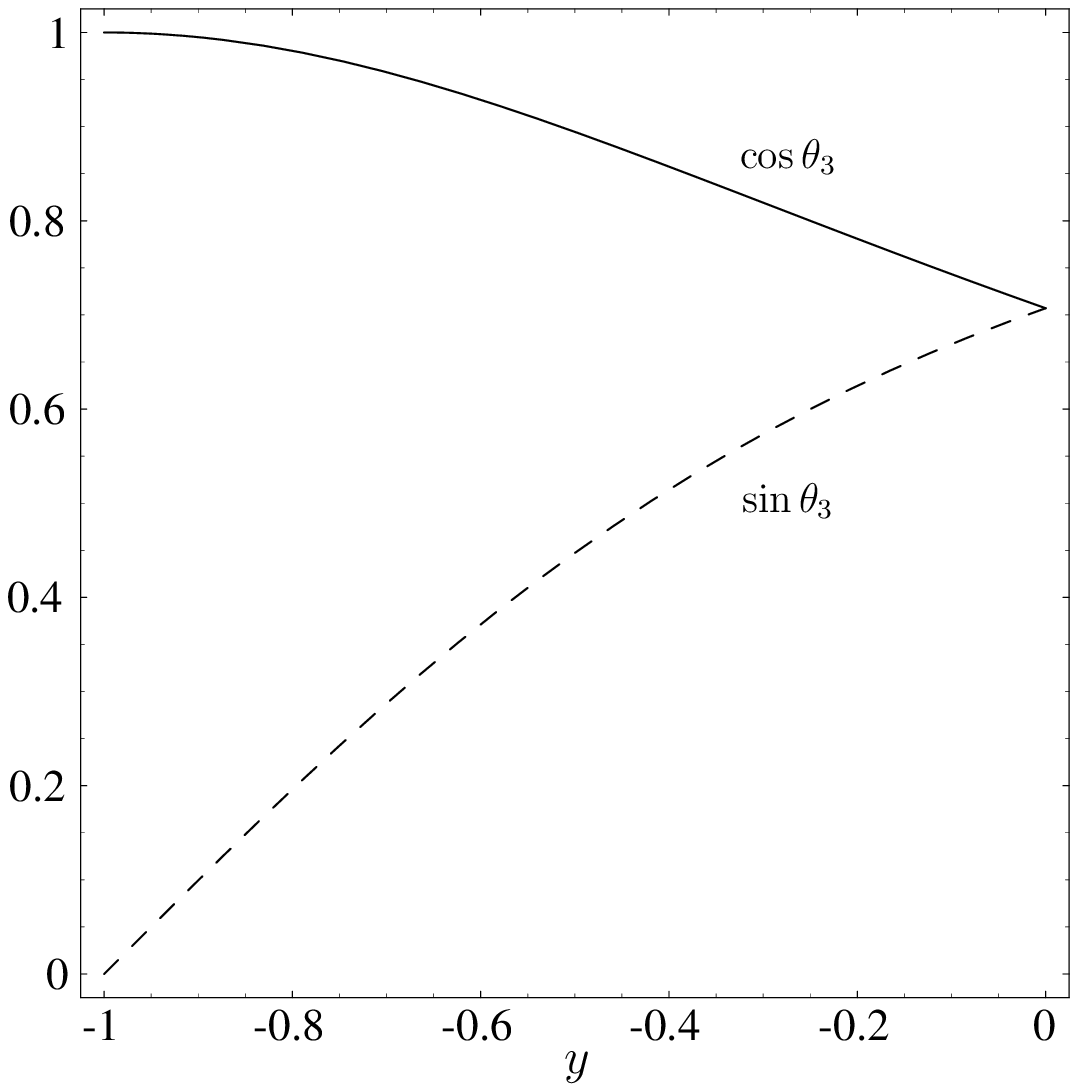, height=6.4cm}\\[-4em]
\hspace{5mm} (a) \hspace{11.7cm} (b)
\vspace{1.5em}
\caption{Illustration of the smooth transition from Majorana ($y=-1$) to Dirac
($y=0$) gluino masses (a) and mixing angles (b).}
\label{fig:pathy}
\end{figure}

The path from the N=1 Majorana theory to the Dirac theory may be defined in
such a way that the mass of the lightest gluino is kept fixed. In addition, we
may identify the off-diagonal mass parameter $M^D_3$ with $m_{{\gw}_1}$ to
reduce the number of free parameters. Starting from the Majorana theory, we
follow the path
\begin{eqnarray} \label{path}
{\mathcal{P}}\;:\;\; M_3 \,\,  &=& m_{{\gw}_1} M'_3 / (M_3' - m_{{\gw}_1})
                                   \quad {\rm for} \quad
                                                 -\infty \leq M_3' \leq 0
                                                   \nonumber \\
                       M^D_3   &=& m_{{\gw}_1}                    \,.
\end{eqnarray}
The heavy gluino mass is trailed along according to
\begin{equation}
  m_{{\gw}_2} = -M'_3 - m^2_{\tilde{g}_1}/(M'_3 - m_{\tilde{g}_1}) \,,
\end{equation}
while the mixing parameters follow from
\begin{equation}
\cos\theta_3 = \frac{1}
 {\sqrt{1+(1-M'_3/m_{\tilde{g}_1})^2}} \,, \qquad
\sin\theta_3  =  \frac{1-M'_3/m_{\tilde{g}_1}}
 {\sqrt{1+(1-M'_3/m_{\tilde{g}_1})^2}} \,. 
\end{equation}
The path ${\mathcal{P}}$ can be mapped onto a unit interval by the
transformation
\begin{equation}
M'_3 = m_{\tilde{g}_1} \, \frac{y}{1+y} \quad {\rm for} \quad
                                            -1 \leq y \leq 0 \,,
\end{equation}
leading to
\begin{equation}
m_{{\gw}_2} = m_{{\gw}_1} \, \biggl (y + \frac{1}{1+y} \biggr ), 
\quad {\rm and} \quad
\cos\theta_3 = \frac{1+y}{\sqrt{1+(1+y)^2}}\,, \quad
\sin\theta_3 = \frac{1}{\sqrt{1+(1+y)^2}}  \,.
\end{equation}
The transition of the parameters $M_3/M'_3$ and $\sin\theta_3/\cos\theta_3$ 
as well as $m_{{\gw}_2}$ is exemplified in Fig.~\ref{fig:pathy}.

For $y=-1$, corresponding to $M'_3 \to -\infty$, the {\underline{Majorana}}
limit for N=1 gluinos is reproduced with the physical mass $m_{{\tilde{g}}_1}$
while the second set of Majorana particles with $m_{{\tilde{g}}_2} \to \infty$
is removed from the system.

In the right-most end-point $y=0$ of the path ${\mathcal{P}}$ the
{\underline{Dirac}} limit is realized. The on-diagonal mass parameters both
vanish while the off-diagonal elements are equal:
\begin{eqnarray}
&& M_3'=M_3 = 0  \;\; {\rm and} \;\;M^D_3 \neq 0  \nonumber \\ &&
\sin\theta_3 = \cos\theta_3 = 1/\sqrt{2}                     \,,
\label{eq:Dirac_case_angle}
\end{eqnarray}
so that the physical masses $m_{{\gw}_{1,2}} = m_{\gw}$ are identical 
and the mixing is maximal.

\subsection{The Dirac Limit}

\noindent
The transformation matrix $\mathcal{U}$, connecting the field bases $\gw/\gw'$
with $\gw_1/\gw_2$, simplifies in the Dirac limit\footnote{In this degenerate
  case the mixing matrix $\mathcal{U}$ is unique up to multiplication on the
  right by an arbitrary orthogonal matrix \cite{CHKZ}.} to
\begin{eqnarray}
\mathcal{U} &=&
         \begin{pmatrix}   \cos\theta_3 & - \sin\theta_3  \\
                        \sin\theta_3 & \cos\theta_3    \end{pmatrix} \cdot
            \begin{pmatrix}  1 & 0  \\
                            0 & i  \end{pmatrix}
  =  \begin{pmatrix}   \cos\theta_3 & -i \sin\theta_3  \\
                       \sin\theta_3 &  i \cos\theta_3  \end{pmatrix} 
\nonumber \\[2mm]
    &\to&   \frac{1}{\sqrt{2}} \begin{pmatrix}   1 & -1  \\
               1 &  1  \end{pmatrix} \cdot
                       \begin{pmatrix}  1 & 0  \\
                                        0 & i  \end{pmatrix}
   = \frac{1}{\sqrt{2}} \begin{pmatrix}   1 & -i  \\
                          1 &  i  \end{pmatrix} \,,
\label{eq:Dirac_mixing}
\end{eqnarray}
corresponding to a $\pi/4$ rotation matrix and a phase matrix which turns the
second eigenvalue positive. Hence the two degenerate physical Majorana fields
$\gw_1$ and $\gw_2$ can be expressed in terms of the original current fields
$\gw'$ and $\gw$ as
\begin{eqnarray}
   \gw_1 &=& \phantom{i}\, [(\gw'_L + \gw'_R) + (\gw_L + \gw_R)]/\sqrt{2}
             \nonumber \\
   \gw_2 &=&  i\,[(\gw'_L - \gw'_R) - (\gw_L - \gw_R)]/\sqrt{2} \,.
\end{eqnarray}
These two Majorana fields are odd and even under charge conjugation:
$\tilde{g}^c_1 = - \tilde{g}^T_1$ and $\tilde{g}^c_2 = + \tilde{g}^T_2$,
respectively.

In this configuration the $t$-channel exchange of the two Majorana fields
$\tilde{g}_{1,2}$ in the processes $q_L q_L' \to \qw_L \qw_L'$ and $q_R q_R'
\to \qw_R \qw_R'$ is maximally destructive and the two amplitudes vanish:
\begin{eqnarray}
   A[q_L q_L' \to \qw_L \qw_L']
      &=& A^*[q_R q_R' \to \qw_R \qw_R']   \nonumber \\
      &\sim&
      \sum_{k=1}^2\, \mathcal{U}_{2k} \mathcal{U}_{2k} = 0   \,.
\end{eqnarray}
The superposition of mixed chiral amplitudes, on the other side, is maximally
constructive:
\begin{equation}
   A[q_L q_R' \to \qw_L \qw_R']
   \sim
   \sum_{k=1}^2\, \mathcal{U}_{2k} \mathcal{U}^\ast_{2k} = 1  \,.
\end{equation}
This picture can be simplified considerably by switching from the two-Majorana
to the Dirac description.

Introducing a superposition of two equal-mass Majorana fields, $m_{\gw_1} =
m_{\gw_2} = m_{\gw}$, the right- and left-handed components of the Dirac
gluino field coupled to the quark current are effectively given by
\begin{eqnarray}
&&   \gw_{D} = (\gw_{1} - i\gw_{2})/\sqrt{2} \;\; : \;\;
     \gw_{DR} = \gw_R          \nonumber \\
&&   {\phantom{\gw_{D} = (\gw_{1} - i\gw_{2})/\sqrt{2} \;\; : \;\;}}\;
     \gw_{DL} = \gw'_L         \,,
     \label{eq:Dirac_field_1}\\[1.5ex]
&&   \gw_{D}^{cT} = -(\gw_{1} + i\gw_{2})/\sqrt{2} \;\; : \;\;
     \gw_{DR}^c = -\gw'^T_R    \nonumber \\
&&   {\phantom{\gw_{D}^{cT} = -(\gw_{1} + i\gw_{2})/\sqrt{2} \;\; : \;\;}}\;
     \gw_{DL}^c = -\gw^T_L     \,.
\label{eq:Dirac_field_2}
\end{eqnarray}
The field ${\gw}_D = (\gw_1-i\gw_2)/\sqrt{2}=\gw_R + \gw'_L$ is a Dirac field,
i.e. it is not self-conjugate: ${\gw}^c_D\neq \pm \gw^T_D$. It describes four
degrees of freedom, the two helicities and the particle/antiparticle
characteristics. The contraction of the field with itself vanishes [in
contrast to Majorana fields], while the contraction between the field and its
conjugate is given by the canonical Dirac value. As a result, this Dirac field
cannot be exchanged between two chirality-L currents and the LL-type amplitude
vanishes. Similarly, the related $\mathcal{C}$-conjugate field $\gw_D^c$ is
coupled to R-type currents but RR amplitudes vanish. On the other hand,
RL-type amplitudes do not vanish and, in fact, the contraction between $\gw_D$
and $\gw_D^c$ generates the usual Dirac propagator, so that the RL amplitude
corresponds to the standard Dirac exchange amplitude. {\it In summa}, the
theory of two mass-degenerate Majorana fields with chiral couplings is
equivalent to the Dirac theory of a single fermion.

The Lagrangian for the super-QCD interaction of gluinos with squarks and
quarks is of the standard N=1 SUSY form (\ref{qcd-yuk}) for one Majorana
gluino mass eigenstate $\tilde{g}$ in the Majorana limit,
but the interaction Lagrangian of the two Majorana gluino fields in the
N=1/N=2 hybrid model can be contracted in the Dirac limit to
\begin{eqnarray}
   \mathcal{L}_{\rm QCD}^{q\tilde{q}\tilde{g}}
       &=& -g_s \frac{1}{\sqrt{2}}
           \left[\, \overline{q_L} \gw_1\, \tilde{q}_L
                  - \overline{q_R} \gw_1\, \tilde{q}_R
                -i (\overline{q_L} \gw_2\, \tilde{q}_L
                  + \overline{q_R} \gw_2\, \tilde{q}_R)
                  +{\rm h.c.} \right] \nonumber\\
       &=& - g_s  \left[\, \overline{q_L} \gw_{D}  \tilde{q}_L
        + \overline{q_R} \gw^{cT}_{D}  \tilde{q}_R + {\rm h.c.} \right]  \,.
\label{eq:dirac_qcd_lagrangian}
\end{eqnarray}
The trilinear gluon/gluino interaction is just the sum of the two individual
standard interactions. The mass term of the Lagrangian in the Majorana limit,
\begin{equation}
   \mathcal{L}^m_{\rm QCD} =
      - \frac{1}{2}
         m_{\tilde{g}} {\rm Tr} [ \bar{\gw} \gw ] \,,
\end{equation}
is altered in the Dirac limit to
\begin{eqnarray}
 \mathcal{L}^m_{\rm QCD}
   &=& - \frac{1}{2}
          m_{\tilde{g}} {\rm Tr}\, [{\bar{\gw}}_{1} \gw_{1}
                                           + {\bar{\gw}}_{2} \gw_{2} ]
 \nonumber \\
   &=& - m_{\tilde{g}}\, {\rm Tr}\, [\overline{\tilde{g}}_D \, \gw_D ]
\end{eqnarray}
in terms of the two degenerate Majorana mass eigenstates, $\gw_1, \gw_2$, and
the Dirac field $\gw_D$ or $\gw_D^c$, respectively.

As will be demonstrated later in several examples, the transition from the
Lagrangian of the 2-Majorana theory to the Dirac theory entails the
isomorphism of the two theories in all dynamical aspects, including the
(properly defined sets of) cross sections.

In the Dirac theory, a conserved quantum number $D$, associated with the
R-symmetry of the N=2 theory noted in Ref.~\cite{no}, can be assigned to each
supersymmetric particle state appearing in the Lagrangian, {\it nota bene} the
interaction term (\ref{eq:dirac_qcd_lagrangian}):
\begin{eqnarray} \label{dc}
&& D[{\tilde{q}}_L] = D[{\tilde{g}}^c_D] = D[\tilde{l}^-_L]
  = D[{\tilde{\chi}}^{c0}_D] = D[{\tilde{\chi}^+_{D1}}]= -1 \,,\nonumber \\
&& D[{\tilde{q}}_R] = D[{\tilde{g}}_D] = D[\tilde{l}^-_R]
  = D[{\tilde{\chi}}^0_D] = D[{\tilde{\chi}}^+_{D2}] = +1\,.
\end{eqnarray}
The ${\tilde{\chi}}^\pm_D$ components relevant for the gauge-strength Yukawa
interactions are $\tilde{\chi}^-_{D1R} = \tilde{W}^-_R$ and
$\tilde{\chi}^+_{D2R} = \tilde{W}^+_R$. Antiparticles carry the Dirac
charges $-D$ correspondingly. The Dirac charge of all SM particles vanishes.
Note that the superpartners of left- and right-handed SM fermions carry
opposite Dirac charge; this implies that terms mixing these fields will not
conserve $D$.  [Electroweak neutralinos and sleptons will be discussed in more
detail in section~\ref{sc:ew}.] The Dirac charge $D$ conveniently classifies
possible production processes and decay modes for the supersymmetric particles
in the Dirac theory, as widely applied in the next sections.\footnote{We could
  equivalently define SM matter fermions to carry non-vanishing $D$, with
  $D[q_L] = -D[q_R]$, with sfermions having vanishing Dirac charge.}

\begin{figure}
\epsfig{figure=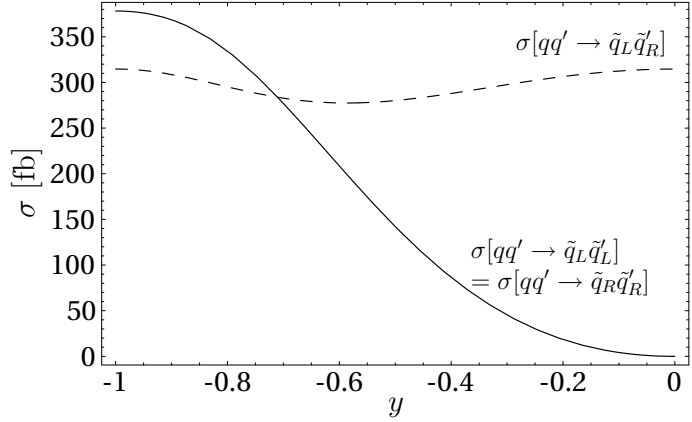, width=9cm}
\vspace{-2ex}
\caption{Partonic cross sections for different-flavor squark production as a
function of the Dirac/Majorana control parameter~$y$. The plot corresponds to a
fixed partonic center-of-mass energy $\sqrt{s}=2000$ GeV, and
$m_{\qw} = 500$ GeV and $m_{\gw_1} = 600$ GeV. The scale of the running
coupling $\alpha_s^{\overline{\mbox{\tiny MS}}}(\mu)$ has been chosen to be
$\mu=\sqrt{s}$.}
\label{fig:sqsQ}
\end{figure}

The cross sections for the processes $qq' \to \qw \qw'$ are characteristically
different in the two limits [for simplicity, we again take equal masses for
$\tilde q_L$ and $\tilde q_R$]:
\begin{eqnarray}
\label{eq:xsec_diff_fl1}
\mbox{Majorana :\ \ } \sigma[qq' \to \qw_L \qw_L']
    &=& \sigma[qq' \to \qw_R \qw_R']     = \frac{2 \pi \alpha_s^2}{9} \,
  \frac{\beta m_{\gw_1}^2}{s m_{\gw_1}^2 + (m_{\gw_1}^2 - m_{\tilde{q}}^2)^2}
 \\[1ex]
\label{eq:xsec_diff_fl2}
\mbox{Dirac :\ \ }
 \sigma[qq' \to \qw_L \qw_L'] &=& \sigma[qq' \to \qw_R \qw_R'] = 0   \\[2ex]
\label{eq:xsec_diff_fl3}
\mbox{Majorana = Dirac :\ \ } \sigma[qq' \to \qw_L \qw_R'] &=&
                        \frac{2 \pi \alpha_s^2}{9s^2} \bigl [
                 (s+2(m_{\gw_1}^2 - \mqw^2)) L_1 - 2 \beta s \bigr]\,,
\end{eqnarray}
where $L_1$ has been defined in Eq.(\ref{eq:xsec_diff_fl}). The cross
sections in the evolution from the Majorana limit to the Dirac limit are
displayed in Fig.~\ref{fig:sqsQ} at the parton level. While the $\qw_L \qw_L'$
cross section moves monotonically to zero, the $\qw_L \qw_R'$ cross section is
only slightly modulated on the path $\mathcal{P}$ from the N=1 Majorana
limit to the Dirac limit. It should be noted that the Dirac cross sections are
identical to the 2-Majorana cross sections owing to destructive interferences
between the $\gw_1$ and $\gw_2$ exchange diagrams.

\paragraph*{\underline{Near-Dirac field:}} Generalizing the definitions
Eqs.~\eqref{eq:Dirac_field_1} and \eqref{eq:Dirac_field_2} for the Dirac field,
the continuous approach of two nearly mass degenerate, nearly chirally coupled
Majorana fields to the Dirac limit can be framed quantitatively. We define the
fields
\begin{eqnarray}
\gw_{\delta } &=& \phantom{-}\sin\theta_3\; {\gw}_{1} - i  \cos\theta_3\; 
  {\gw}_{2}
 \equiv \cos\delta \; {\gw}_{D} + \sin\delta\; {\gw}^{cT}_{D}\,,            \\
 \nonumber
\gw^{cT}_{\delta } &=& -\sin\theta_3\; {\gw}_1 - i \cos\theta_3\; {\gw}_2
 \equiv \sin\delta \; {\gw}_{D} + \cos\delta\; {\gw}^{cT}_{D}               \,.
\end{eqnarray}
With $\cos/\sin\delta = (\cos\theta_3 \pm \sin\theta_3)/\sqrt{2}$ they are
identical to the standard Dirac fields for $\cos\theta_3 = \sin\theta_3 =
1/\sqrt{2}$, or $\delta=0$, but keep the approximate character of Dirac fields
nearby ($0 < |\delta| \ll 1$). The contraction of the field with itself, $\sim
\cos 2 \theta_3$, nearly vanishes for $\theta_3 \sim \pi/4$, while the
contraction with the conjugate fields is unity.  Thus the LL and RR transition
amplitudes are proportional to $\cos2\theta_3$ and non-zero, while the LR
transition remains 1.  The exchanges of the near-Dirac fields is equivalent to
the exchanges of the two Majorana fields, generating transition amplitudes LL,
RR $= \cos^2\theta_3 - \sin^2\theta_3 = \cos 2\theta_3$ and LR
$=\cos^2\theta_3 + \sin^2\theta_3 = 1$.  Taking the N=2 gluino as an example
in the limit $M_3' \to 0$, the parameters describing the approach to the Dirac
field are given by
\begin{eqnarray}
   \cos\theta_3  &=& \frac{1}{\sqrt{1+(1-M'_3/m_{\tilde{g}_1})^2}} \approx
                  \frac{1}{\sqrt{2}} \left[ 1 + \frac{1}{2} \frac{M'_3}{m_{\tilde{g}_1}}\right] \\
   \sin\theta_3  &=& \frac{1 -M'_3/m_{\tilde{g}_1}}{
                  \sqrt{1+(1-M'_3/m_{\tilde{g}_1})^2}}  \approx
                  \frac{1}{\sqrt{2}} \left[ 1 - \frac{1}{2} \frac{ M'_3}{m_{\tilde{g}_1}} \right]
          \,,
\end{eqnarray}
generating
\begin{equation}
\cos\delta \approx 1 \qquad {\rm and} \qquad
\sin\delta \approx M'_3/2m_{\tilde{g}_1}.
\end{equation}
In contrast to the wave functions, the two mass eigenvalues
$m_{{\tilde{g}}_{1,2}}$ remain equal up to second order in $M'_3$. As a
result, exchanging the near-Dirac fields between L- and R-currents reproduces
the cross sections calculated otherwise by the exchange of the almost
degenerate Majorana fields.

\subsection{Summary of Characteristic Scattering Processes}

\noindent
The entire ensemble of partonic cross sections for the N=1 Majorana theory has
been calculated in Ref.~\cite{Beenakker}, improving on the Born approximations
\cite{LlSmith} by including the radiative super-QCD corrections [for threshold
resummations see \cite{Kul}]. Electroweak tree-level contributions to the
production of two (anti)squarks have been calculated in Ref.~\cite{bddk},
while electroweak one-loop corrections to squark antisquark production have
been derived in Ref.~\cite{hollik}. Since the number of reactions is
approximately tripled when the theory is followed along the Majorana-Dirac
path, we restrict the discussion to a set of characteristic
examples.\footnote{The complete set of cross sections is available at
    {\tt http://www.pitt.edu/\~{}afreitas/formulas.pdf}.} To highlight the
  characteristic differences between Majorana and Dirac theories, it is
  sufficient to work out the cross sections at the Born level.

\paragraph*{(a) \underline{Different-flavor quark scattering:}}\

These channels have been used in the previous sections to develop the
differences between Majorana and Dirac theories. The results are presented
in Eqs.~\eqref{eq:xsec_diff_fl},
\eqref{eq:xsec_diff_fl1}--\eqref{eq:xsec_diff_fl3} and Fig.~\ref{fig:sqsQ}.

\paragraph*{(b) \underline{Different-flavor quark-antiquark scattering:}}\

\begin{figure}
\psfig{figure=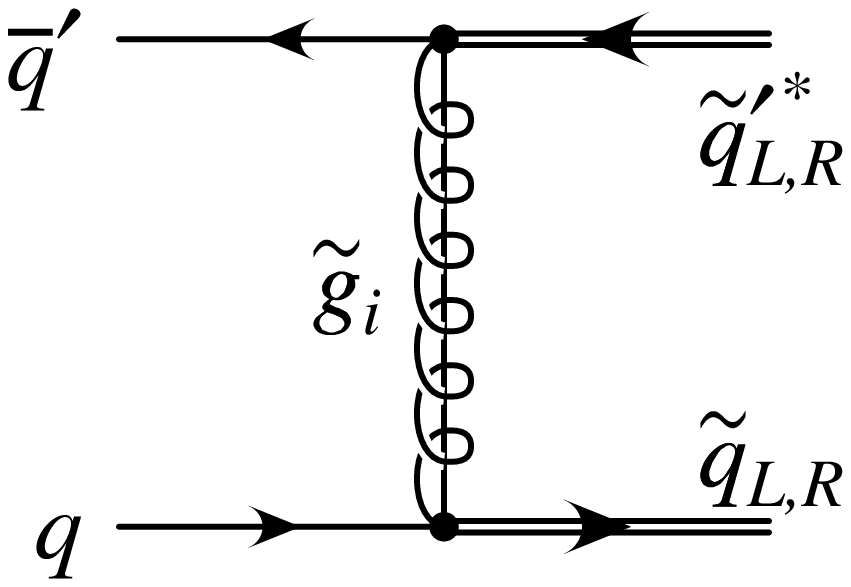, width=4.5cm}
\hspace{1cm}
\psfig{figure=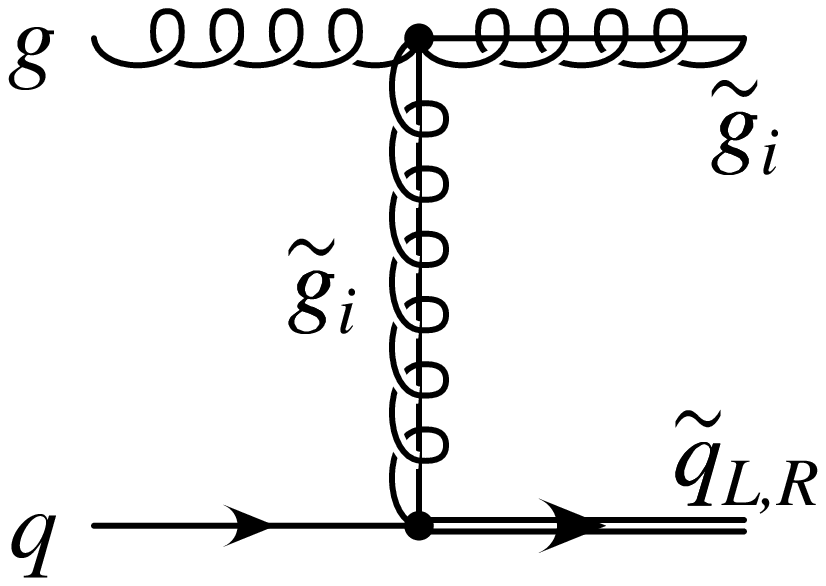, width=3.5cm}
\psfig{figure=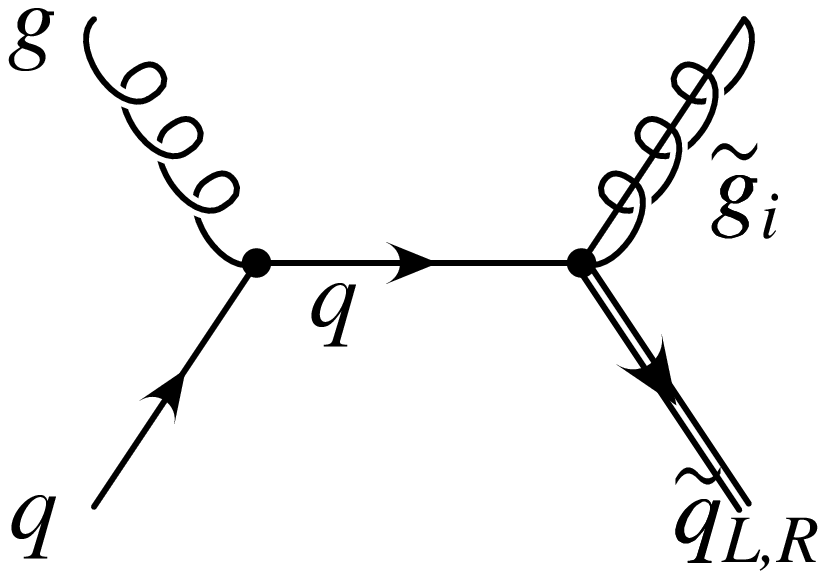, width=3.5cm}
\\[-2ex]
\makebox[4.5cm]{(a)}
\hspace{1cm}
\makebox[3.5cm]{(b)}
\makebox[3.5cm]{(c)}
\\[3ex]
\psfig{figure=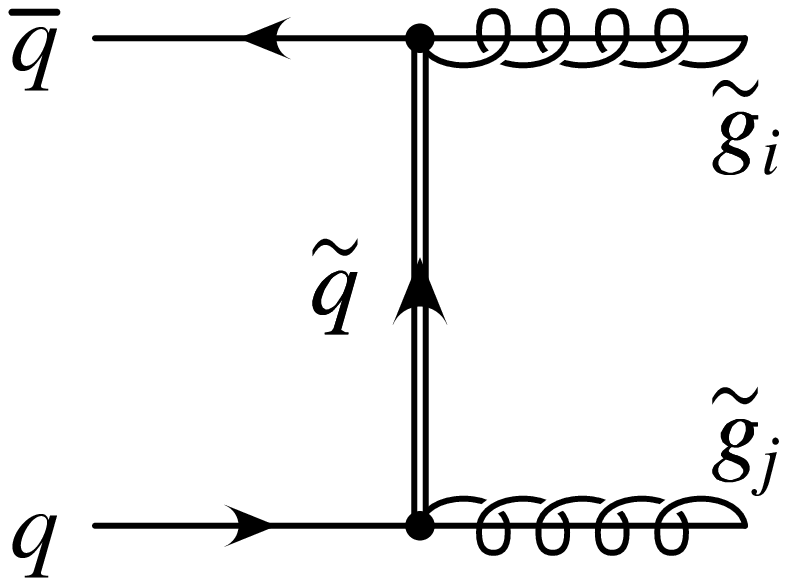, width=3.5cm}
\psfig{figure=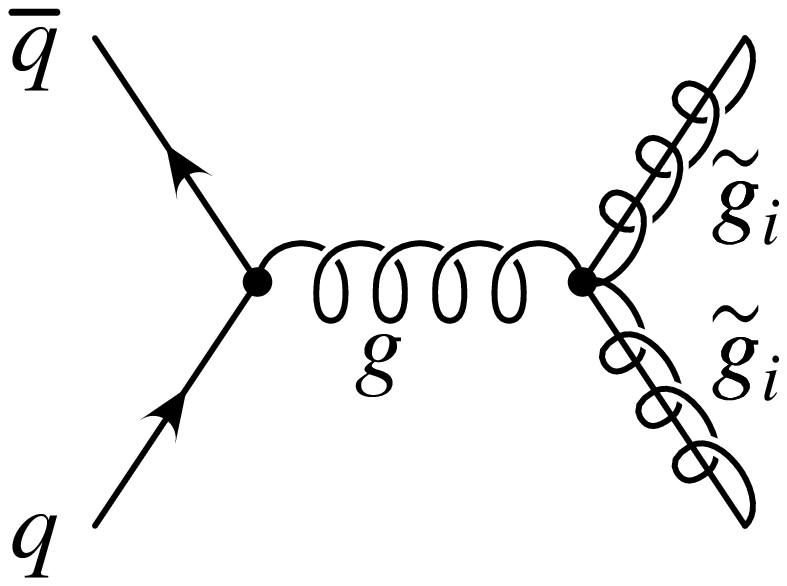, width=3.5cm}
\hspace{1cm}
\psfig{figure=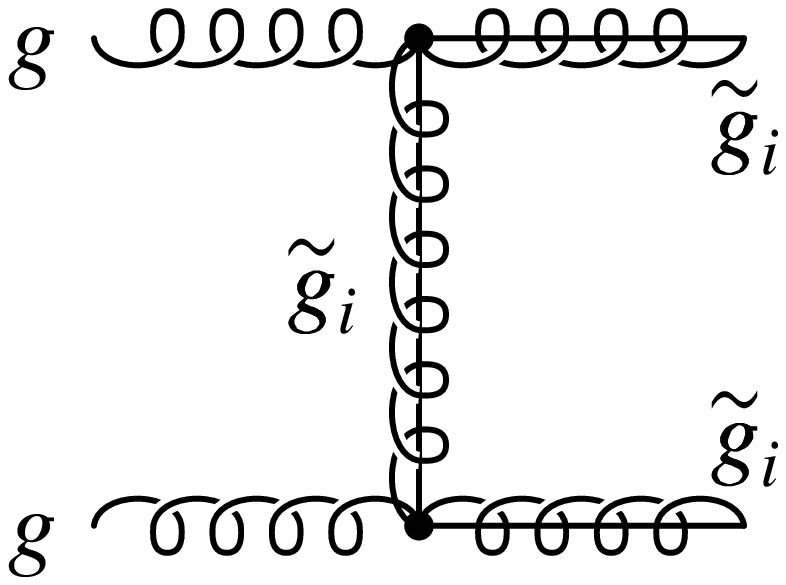, width=3.2cm}
\psfig{figure=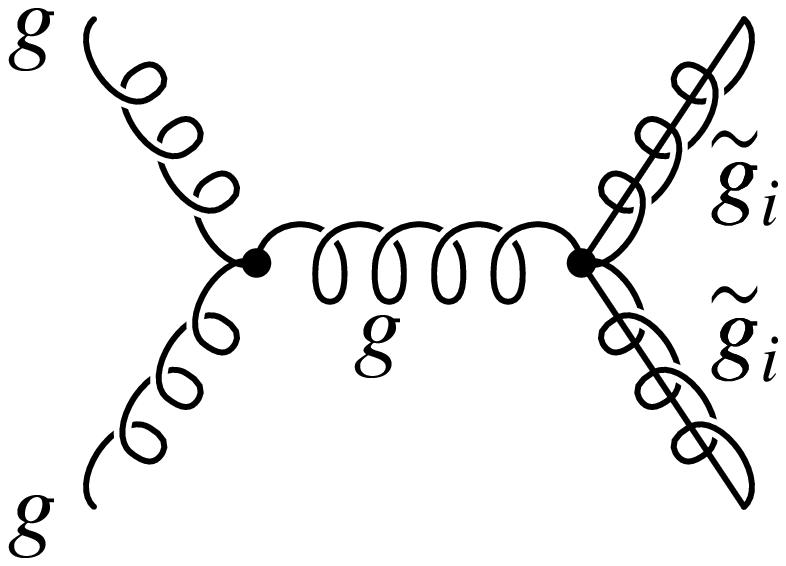, width=3.2cm}
\\[-2ex]
\makebox[3.5cm]{(d)}
\makebox[3.5cm]{(e)}
\hspace{1cm}
\makebox[3.2cm]{(f)}
\makebox[3.2cm]{(g)}
\caption{Feynman diagrams for different-flavor squark production in quark
annihilation (a), squark-gluino production (b,c), gluino production in quark
annihilation (d,e) and gluino production in gluon fusion (f,g). [The indices
$i,j$ count the two gluinos in the N=2 hybrid model, and should be ignored
for the N=1 MSSM.]} 
\label{fig:dia2}
\end{figure}

\begin{figure}
\epsfig{figure=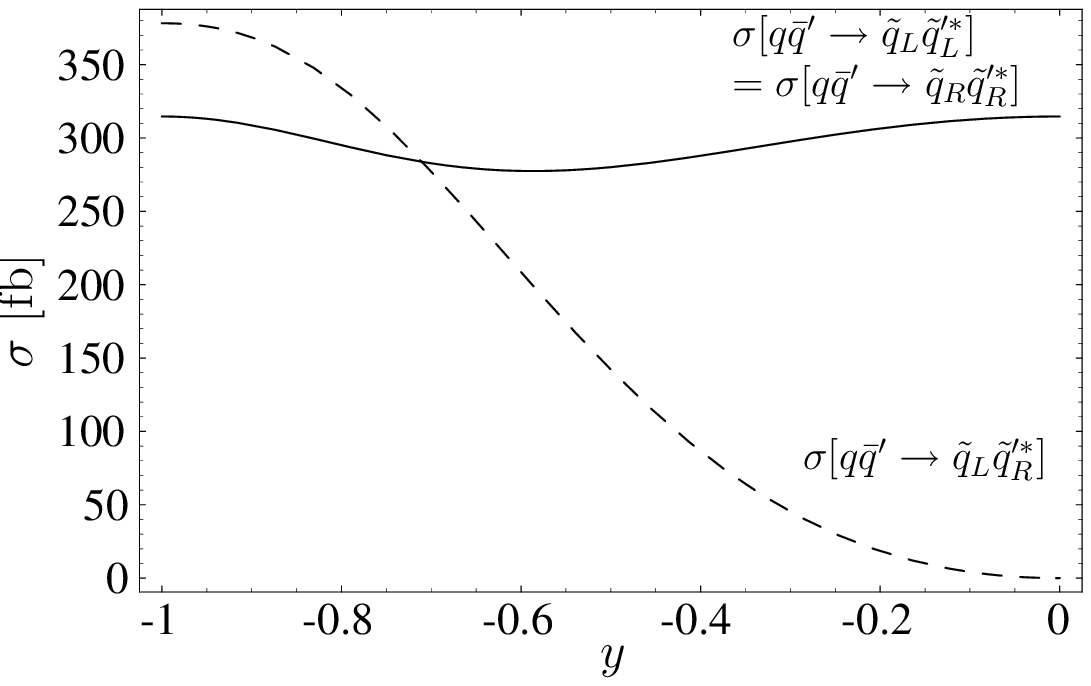, height=5cm}
\hspace{1cm}
\epsfig{figure=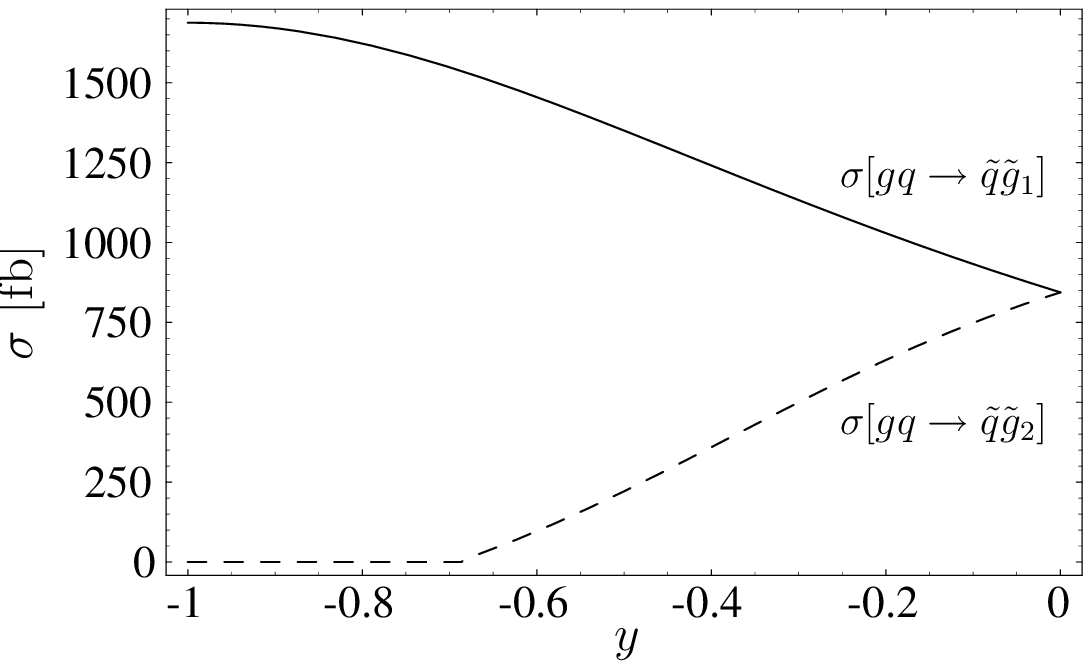, height=5cm}
\\[-2em]
(a) \hspace{9cm} (b) \hfill ~
\\[1em]
\epsfig{figure=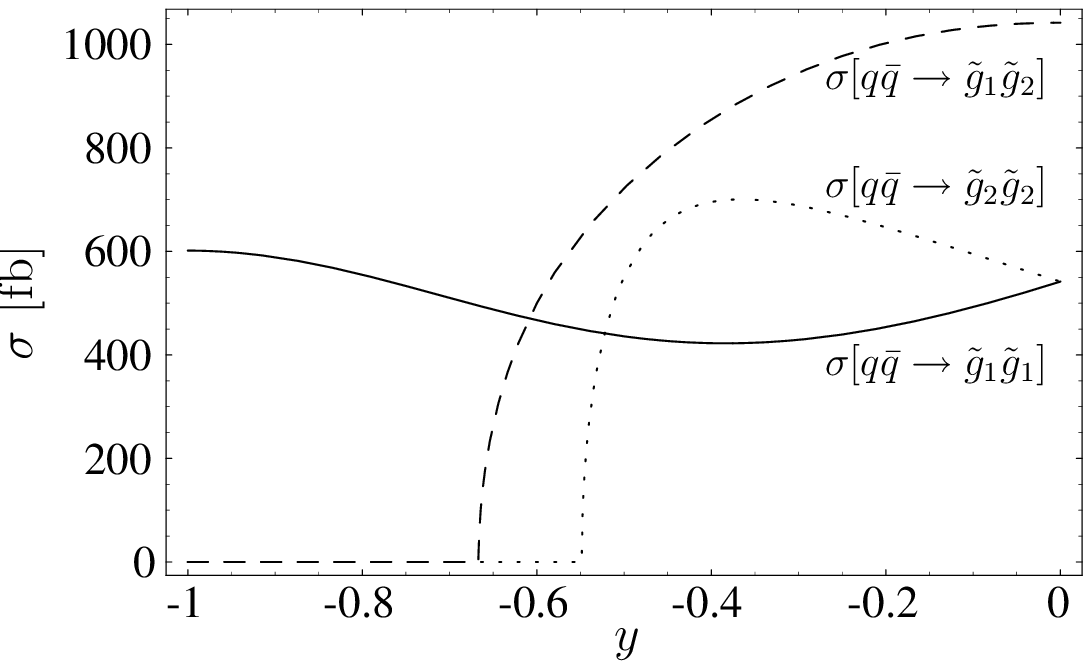, height=5cm}
\hspace{1cm}
\epsfig{figure=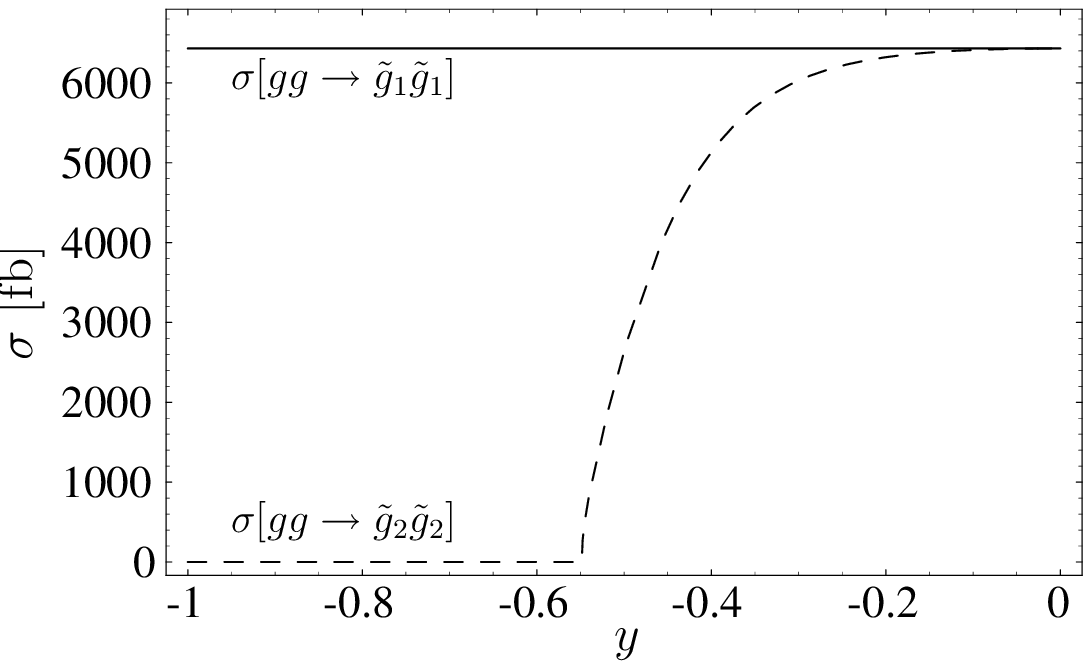, height=5cm}
\\[-2em]
(c) \hspace{9cm} (d) \hfill ~
\caption{Partonic cross sections for different-flavor squark production in
quark annihilation (a), squark-gluino production (b) and gluino production
(c,d). The cross sections are given as functions of the Dirac/Majorana control
parameter~$y$. Soft breaking parameters are as in Fig.~\ref{fig:sqsQ}.}
\label{fig:ploty}
\end{figure}

The Feynman diagrams for $q \bar{q}' \to \qw_L \qw'^\ast_L$, $\qw_R
\qw'^\ast_R$, $\qw_L \qw'^\ast_R$ are shown in Fig.$\,$\ref{fig:dia2}~(a).
In the Majorana and Dirac limits, the partonic cross sections read
\begin{eqnarray}
\label{eq:xsec_ann_fl1}
  \mbox{Majorana = Dirac :\ \ } \sigma[q \bar{q}' \to \qw_L \qw_L'^\ast]
    &=& \sigma[q \bar q' \to \qw_R \qw_R'^\ast]
     = \frac{2 \pi \alpha_s^2}{9s^2} \bigl [
                 (s+2(m_{\gw_1}^2 - \mqw^2)) L_1 - 2 \beta s \bigr]
\\[2ex]
\label{eq:xsec_ann_fl2}
\mbox{Majorana :\ \ } \sigma[q \bar{q}' \to \qw_L \qw_R'^\ast] &=&
     \frac{2 \pi \alpha_s^2}{9} \,
   \frac{ \beta m_{\gw_1}^2}{s m_{\gw_1}^2 + (m_{\gw_1}^2 - m_{\tilde{q}}^2)^2}
\\[1ex]
\label{eq:xsec_ann_fl3}
  \mbox{Dirac :\ \ }  \sigma[q \bar{q}' \to \qw_L \qw_R'^\ast] &=& 0.
\end{eqnarray}
As before, $\beta = (1-4 \mqw^2/s)^{1/2}$ is the velocity of the produced
squarks. Numerical results for the cross sections along the path $-1\to y\to
0$ are displayed in Fig.~\ref{fig:ploty}~(a).

For equal-flavor quark-antiquark scattering the additional gluino $s$-channel
exchange must be added to the $t$-channel exchange diagrams.

\paragraph*{(c) \underline{Squark-gluino production:}}\

The Feynman diagrams for the super/hyper-Compton processes $gq \to \qw \gw_1$,
$\qw \gw_2$ are given in Fig.~\ref{fig:dia2}~(b,c). As before, we give
formulas for the cross sections in the two limiting cases:
\begin{eqnarray}
\label{eq:xsec_comp1}
  \mbox{Majorana = Dirac :\ \ } \sigma[gq \to \qw_{L,R} \gw]
    &=& \sigma[gq \to \qw_L \gw_D] = \sigma[gq \to \qw_R \gw^c_D] \\
    &=& \frac{\pi \alpha_s^2}{18s^3} \bigl [
                 2(4s-4m_{\gw_1}^2 - 5\mqw^2)(m_{\gw_1}^2 - \mqw^2) L'_1
\nonumber \\ && \hspace{2em}
         +9(s(s+2m_{\gw_1}^2)+2\mqw^2(\mqw^2-m_{\gw_1}^2-s)) L_1
\nonumber \\ && \hspace{2em}
         - \beta s (7s + 32(m_{\gw_1}^2 - \mqw^2))\bigr]
\nonumber \\[1ex]
\label{eq:xsec_comp2}
  \mbox{Dirac :\ \ }  \sigma[gq \to \qw_L \gw^c_D] &=&
    \sigma[gq \to \qw_R  \gw_D] = 0,
\end{eqnarray}
with
\begin{equation}
L_1 = \log\frac{(1+\beta)+2( m_{\tilde{g}_1}^2 - 2 m_{\tilde{q}}^2)/s}
         {(1-\beta)+2( m_{\tilde{g}_1}^2 - 2 m_{\tilde{q}}^2)/s}
\qquad\qquad
L'_1 = \log\frac{(1+\beta)-2( m_{\tilde{g}_1}^2 - 2 m_{\tilde{q}}^2)/s}
         {(1-\beta)-2( m_{\tilde{g}_1}^2 - 2 m_{\tilde{q}}^2)/s}.
\label{eq:l1}	 
\end{equation}
Here $\beta = \left([s-(\mqw + m_{\gw_1})^2][s-(\mqw - m_{\gw_1})^2]
\right)^{1/2} /s$ denotes the momenta of the final-state squarks and gluinos
in units of half the total c.m. parton energy, i.e. the velocity for
equal-mass particles.  Fig.~\ref{fig:ploty}~(b) shows the cross sections for
the two Majorana mass eigenstates along the interpolated path between the two
limits. As can be seen in the figure, the second gluino $\gw_2$ can only be
produced if, for the fixed parton c.m.~energy, it becomes light enough so that
the kinematical threshold is crossed. Approaching the Dirac limit $y\to 0$,
the cross sections for $\qw \gw_1$ and $\qw \gw_2$ production become equal.
Note that the total $\tilde q \tilde g$ production cross section is the same
in the Dirac limit as in the original MSSM.

\paragraph*{(d) \underline{Gluino pairs:}}\

Gluino pairs can be produced through quark annihilation, $q \bar{q} \to \gw_1
\gw_1$, $\gw_1 \gw_2$, $\gw_2 \gw_2$ (see Fig.~\ref{fig:dia2}~(d,e)), or
through gluon fusion $gg \to \gw_1 \gw_1$, $\gw_2 \gw_2$ (see
Fig.~\ref{fig:dia2}~(f,g)). The production of gluino pairs in gluon-gluon
collisions is based solely on QCD gauge interactions. By conservation of the
color current, mixed $\gw_1,\gw_2$ gluino pair production is therefore not
possible. The cross sections are given by
\begin{eqnarray}
\label{eq:xsec_glgl1}
  \mbox{Majorana :\ \ } \sigma[q \bar{q} \to \gw \gw]
    &=& \frac{4\pi \alpha_s^2}{27s^3} \biggl [
        -   2 \frac{8 s^2 \, \mqw^2
            + s(7 m_{\gw_1}^4 - 32 m_{\gw_1}^2\, \mqw^2 + 25\mqw^4)
            - 18 (m_{\gw_1}^2 - \mqw^2)^3}
               {s-2 m_{\gw_1}^2 + 2 \mqw^2} \; L_1'
\nonumber \\ && \hspace{3em}
        + \beta \biggl (13 s^2 - 6 s (m_{\gw_1}^2-3\mqw^2) -
            \frac{8 s^3 \mqw^2}{s\,\mqw^2 + (m_{\gw_1}^2-\mqw^2)^2}
        \biggr ) \biggr ]
\end{eqnarray}
\begin{eqnarray}
\label{eq:xsec_glgl2}
  \mbox{Dirac :\ \ }  \sigma[q \bar{q} \to \gw_D \gw_D^c]
    &=& \frac{4\pi \alpha_s^2}{27s^3} \biggl [
        -   2 \bigl (9 (m_{\gw_1}^2 - \mqw^2)^2 + s(m_{\gw_1}^2
            + 8\mqw^2) \bigr ) L_1'
\nonumber \\ && \hspace{3em}
        + \beta \biggl (19 s^2 + 6 s (m_{\gw_1}^2+3\mqw^2) -
            \frac{8 s^3 \mqw^2}{s\,\mqw^2 + (m_{\gw_1}^2-\mqw^2)^2}
        \biggr ) \biggr ]
\\[1ex] \nonumber
    \sigma[q \bar{q} \to \gw_D \gw_D] &=&
    \sigma[q \bar{q} \to \gw_D^c \gw_D^c]  = 0,
\end{eqnarray}
and
\begin{eqnarray}
\label{eq:xsec_glgl3}
  \mbox{Majorana :\ \ } \sigma[gg \to \gw \gw]
    &=& \frac{3\pi \alpha_s^2}{4s^3} \bigl [
            3(s^2+4 s \, m_{\gw_1}^2 -
            4 m_{\gw_1}^4) \log \tfrac{1+\beta}{1-\beta}
        - \beta s (4 s + 17  m_{\gw_1}^2) \bigr ]
\\[2ex]
\label{eq:xsec_glgl4}
  \mbox{Dirac :\ \ }  \sigma[gg \to \gw_D \gw_D^c]
    &=& \frac{3\pi \alpha_s^2}{2s^3} \bigl [
            3(s^2+4 s \, m_{\gw_1}^2 -
            4 m_{\gw_1}^4) \log \tfrac{1+\beta}{1-\beta}
        - \beta s (4 s + 17  m_{\gw_1}^2) \bigr ]
\\ \nonumber
    \sigma[gg \to \gw_D \gw_D] &=& \sigma[gg \to \gw_D^c \gw_D^c]  = 0,
\end{eqnarray}
in the same notation as before,
with $L'_1$ defined in \eqref{eq:l1} but
$\beta = (1-4 m_{\gw_1}^2/s)^{1/2}$.  As the nature of the gluino is changed
smoothly from Majorana to Dirac along the path ${\mathcal{P}}$, several
thresholds are crossed for fixed parton c.m.~energy, see
Fig.~\ref{fig:ploty}~(c,d), whenever the second Majorana particle becomes
light enough to allow $\gw_1\gw_2$ and $\gw_2\gw_2$ pair production,
respectively.

Again, the identity of the 2-Majorana with the Dirac theory can be
re-examined by verifying the equality of the cross sections,
\begin{eqnarray}
    \sigma[q \bar{q} \to \gw_D \gw_D^c] &=&
    \sum^2_{k,l=1} \sigma[q \bar{q} \to \gw_k \gw_l]\,,  \nonumber \\
    \sigma[gg \to \gw_D \gw_D^c] &=&
    \sum^2_{k} \sigma[gg \to \gw_k \gw_k] \,,
\end{eqnarray}
in a meticulous accounting of interference effects in double-gluino production. 
In the Dirac limit, the total production cross section for
$gg \rightarrow gluinos$ is therefore twice as large as in the MSSM.

The hadron cross sections will be discussed for the LHC in the final section,
including crucial tests for discriminating the MSSM Majorana theory from a
Dirac theory experimentally.

\subsection{Gluino Decays}

\noindent
If squarks are heavier than gluinos the dominant channels are decays to
gluinos. Otherwise squarks decay into electroweak chargino and neutralino
channels. Gluinos in turn always decay to pairs of quarks and squarks, either
real or virtual.  The partial widths of all these strong \cite{Beenakker2} and
electroweak modes \cite{Djouadi} are known in next-to-leading order in N=1
supersymmetry. In this subsection only strong decay channels will be
discussed, while electroweak decays are postponed to the next section.

{(a) $\mqw > \mgw$:} The partial widths for squark decays to Majorana and
Dirac gluinos,
\begin{eqnarray}
     \Gamma [\qw_L \to q \gw ]
   = \Gamma [\qw_R \to q \gw ]
   = \Gamma [\qw_L \to q \gw^c_D ]
   &=& \Gamma [\qw_R \to q \gw_D ]
   =  \frac{2 \alpha_s}{3} \frac{(m^2_{\qw} - m^2_{\gw})^2}{m^3_{\qw}} \,, \\
     \Gamma [\qw_R \to q \gw^c_D ] &=& \Gamma [\qw_L \to q \gw_D ] = 0 \,,
\end{eqnarray}
[and correspondingly for the charge-conjugate states] are the same for equal
masses and couplings. This applies for the two endpoints of the path
$\mathcal{P}$, the standard N=1 Majorana limit and the Dirac limit.  Even
though the decay mechanism is strong, the P-wave decay width is suppressed
nevertheless when the squark/gluino mass difference becomes small.

{(b) $\mgw > \mqw$:} A similar relation applies for Majorana and Dirac
gluino decays into quarks and squarks [and charge-conjugate states]:
\begin{eqnarray}
   \Gamma [\gw \to q \qw^*_L ] = \Gamma [\gw \to q \qw^*_R ]
                   = \Gamma [\gw_D \to q \qw^*_L ]
                           = \Gamma [\gw_D^c \to q \qw^*_R ]
                           &=& \frac{\alpha_s}{8}
                             \frac{(m^2_{\gw} - m^2_{\qw})^2}{m^3_{\gw}} \,,
\label{eq:sqdec1}
\\
  \Gamma [\gw_D \to q \qw^*_R ]
                           = \Gamma [\gw_D^c \to q \qw^*_L ] &=& 0\,.
\label{eq:sqdec2}
\end{eqnarray}
Non-isotropic angular distributions follow the familiar $\cos\theta$
distribution for spin 1/2 $\to$ spin 1/2 $+$ spin 0 decays.

Squarks decay either to Dirac particles or to Dirac antiparticles. These
modes can be distinguished in the subsequent Dirac decays.

The Majorana or Dirac character of the gluinos can be demonstrated nicely in
the charge assignments of squarks in the decays of gluino pairs:
\begin{eqnarray}
 {\rm Majorana}: \quad \phantom{_D}
            \gw           &\to& q_L\, {\tilde{q}}^\ast_L \oplus
                    \overline{q_L}\, \tilde{q}_L \ \ \mbox{and}\ \
                    q_R\, {\tilde{q}}^\ast_R \oplus
                    \overline{q_R}\, \tilde{q}_R
\label{eq:gldec1}
\\[1.5ex]
   {\rm Dirac}   : \quad
           \gw_D         &\to& q_L\, {\tilde{q}}^\ast_L \ \ \mbox{and} \ \
                    \overline{q_R} \tilde{q}_R  \nonumber  \\
           \gw^c_D   &\to& \overline{q_L}\, {\tilde{q}}_L \ \ \mbox{and} \ \
                    q_R \tilde{q}^\ast_R ,
\label{eq:gldec2}
\end{eqnarray}
where $\oplus$ connects final states that are produced at equal rates.

For the first two generations, mixing between L- and R-squarks is expected to
be negligible. In this case, the chiralities of the squarks can be
distinguished clearly by their decay modes. For instance, if the lightest
neutralino is mainly bino and the next-to-lightest neutralino is dominantly
wino, the L-squarks have sizable branching fractions into decay cascades
leading to additional leptons, $\tilde{q}_L \to q \, \tilde\chi^0_2 \to
q\,l^+l^-\tilde\chi^0_1$ or $\tilde{q}_L \to q \, \tilde\chi^\pm_1 \to
q\,l^\pm \nu_l\tilde\chi^0_1$, $l = e, \mu, \tau$. On the other hand,
R-squarks would almost always decay directly to the lightest neutralino,
$\tilde{q}_R \to q \, \tilde\chi^0_1$. Furthermore, the decay chain
$\tilde{q}_L \to q \, \tilde\chi^\pm_1 \to q\,l^\pm \nu_l\tilde\chi^0_1$
allows to determine the charge of the $\tilde{q}_L$ experimentally.

Production of Majorana gluino pairs leads to equal amounts of same-sign and
opposite-sign L-squarks, while Dirac gluino pairs generate only the ordinary
opposite-sign combination:
\begin{eqnarray}
   {\rm Majorana}: \qquad\:\,  pp \to \gw \gw
      &\to& q\, q\, {\tilde{q}}^\ast_L\, {\tilde{q}}^\ast_L           \oplus
    q\, \bar{q}\, {\tilde{q}}_L\, {\tilde{q}}^\ast_L \oplus
    \bar{q}\, q\, {\tilde{q}}^\ast_L\, {\tilde{q}}_L\oplus
    \bar{q}\, \bar{q}\, {\tilde{q}}_L\, {\tilde{q}}_L
 \\[1.5ex]
   {\rm Dirac}: \quad    pp \to \gw_D \gw_D     &=& 0     \nonumber \\
            pp \to \gw_D \gw_D^c
      &\to&
    q\, \bar{q}\, {\tilde{q}}_L\, {\tilde{q}}^\ast_L      \,,
\end{eqnarray}
and correspondingly for R-squarks and mixed L/R final states.  In section
\ref{sec:lhc_phenomenology} the LHC phenomenology of this process will be
discussed in more detail.

For gluino decays into tops and stops the situation is more complex due to
potentially sizable stop mixing. Nevertheless, unless the stop mixing is
maximal, i.e. $\theta_{\tilde{t}} = \pi/4$, Dirac gluinos will lead to
an asymmetry in the stop charge assignment as a result of mass difference
between the two stop mass eigenstates:
\begin{eqnarray}
   {\rm Majorana}:  \qquad\:\, pp \to \gw \gw
      &\to&
    t\, t\, {\tilde{t}}^\ast\, {\tilde{t}}^\ast           \oplus
    t\, \bar{t}\, {\tilde{t}}\, {\tilde{t}}^\ast \oplus
    \bar{t}\, t\, {\tilde{t}}^\ast\, {\tilde{t}}\oplus
    \bar{t}\, \bar{t}\, {\tilde{t}}\, {\tilde{t}}
 \\[1.5ex]
\label{diractop}
   {\rm Dirac}: \quad    pp \to \gw_D \gw_D     &=& 0     \nonumber \\
                   pp \to \gw_D \gw_D^c
                    &\to&
    \alpha_D  \left( t_L\, t_R\, {\tilde{t}}^\ast\, {\tilde{t}}^\ast    \oplus
    \overline{t_L}\, \overline{t_R}\, {\tilde{t}}\, {\tilde{t}} \right)\ \
                   \mbox{and} \ \ 
    \beta_D  \left( t_L\, \overline{t_L}\, {\tilde{t}}\, {\tilde{t}}^\ast
                   \oplus 
     \overline{t_R}\, t_R\, {\tilde{t}}^\ast\, {\tilde{t}} \right)\, .
\end{eqnarray}
The gluinos will decay with a larger branching fraction into the lighter of
the two stop states. For Majorana pairs this leads to universal charge
assignments independent of stop mixing. On the other hand, a Dirac gluino
$\gw_D$ (antigluino $\gw^c_D$) decays more often into a stop (antistop) if
the lighter stop state is mostly R-chiral. If the lighter stop is mostly
L-chiral, the opposite decay patterns dominate. Either way one obtains
\begin{equation}
\alpha_D \, < \, \beta_D \,,
\end{equation}
leading to more opposite-sign top pairs than same-sign top pairs in the final
state. In addition, the Majorana gluinos $\gw$ decay to top and antitop quarks
of both chiralities L,R with equal probability while the Dirac gluino pairs
$\gw_D \gw^c_D$ decay to quarks [or antiquarks] which carry different L and R
chiralities as indicated in Eq.(\ref{diractop}), giving rise to different
decay distributions.

In the Dirac theory the ``Majorana-like'' decay pattern $\alpha_D =
\beta_D$ can only be realized for maximal stop mixing. Using leptonic decay
modes of the top quarks to identify their charge, multi-top final states
therefore offer a powerful testing ground for distinguishing Majorana from
Dirac gluinos.

\section{The Electroweak Sector}
\label{sc:ew}

\noindent
If for N=1 the supersymmetry breaking scale is much larger than the
electroweak mass scale $v$, the neutralino sector includes two Majorana
gauginos associated with the hypercharge U(1) and the isospin SU(2) gauge
groups, and two nearly mass degenerate Majorana higgsinos. Thus, in the limit
$v/\mu \to 0$ the system consists of two Majorana gauginos and one Dirac
higgsino.  Extending the N=1 supersymmetry to the N=2 supersymmetry, the two
gaugino degrees of freedom are doubled and, in parallel to the gluino sector,
the two U(1) and SU(2) related gaugino fields may transform from Majorana to
Dirac fields.{\footnote{The discussion of the electroweak sector is
    restricted, almost exclusively, to those points which affect the
    phenomenology of squark/gluino decays; the only exception will be
    selectron pair production for polarized beams.}}

\subsection{N=1/N=2 Neutralino and Chargino Masses and Spinor Wave Functions}

\noindent
In the limit of asymptotically high N=2 supersymmetry scales, the neutralino
mass matrix \eqref{eq:neu_mass_matrix} disintegrates into three weakly coupled
2$\times$2 sub-matrices associated with the gauginos of the gauge groups U(1)
and SU(2), and the higgsino sector. If the new gaugino mass parameters
$M'_{1,2}$ are infinitely large, the system is reduced to the familiar N=1
MSSM. On the other hand, if the on-diagonal elements of the two 2$\times$2
gaugino sub-matrices vanish and the sub-matrices are reduced to equal
off-diagonal elements, the two Majorana fields of each group can be joined to
a Dirac field. In the limit $v \to 0$ the mechanisms operate strictly parallel
to the gluino sector.

Since the N=1 Majorana limit for {\underline{neutralinos}} has been worked
out in all of its facets in the past, we will here restrict ourselves solely
to the discussion of the Dirac/near-Dirac limit. The original current fields 
in Cartesian coordinates are denoted by
\begin{equation}
   \tilde\chi_{\rm curr}
     = \{\tilde{B}', \tilde{B}, \tilde{W}'^3, \tilde{W}^3,
                         \tilde{H}_d, \tilde{H}_u \}^T
\end{equation}
the mass eigenfields, for $v \to 0$, are maximally mixed superpositions of the
current eigenfields:
\begin{eqnarray}
   \tilde\chi_{\rm mass} &=& \{\tilde{B}_1, \tilde{B}_2,
                           \tilde{W}^3_1, \tilde{W}^3_2,
                           \tilde{H}_1, \tilde{H}_2\}^T\,,
\label{eq:Majorana_neutralino_fields}
\end{eqnarray}
where, for real and non-negative $M^D_1, M^D_2$ and $\mu$, the six mass
eigenstates are written in terms of the current fields as
\begin{eqnarray}
&& \tilde{B}_{1,2} = \{i\} [(\tilde{B}'_L \pm \tilde{B}'_R)
                         \pm (\tilde{B}_L \pm \tilde{B}_R)]/\sqrt{2}
 \nonumber\\
&& \tilde{W}_{1,2} = \{i\} [(\tilde{W}'_L \pm \tilde{W}'_R)
                         \pm (\tilde{W}_L \pm \tilde{W}_R)]/\sqrt{2}
 \nonumber\\
&& \tilde{H}_{2,1} = \{i\} [(\tilde{H}_{uL} \pm \tilde{H}_{uR})
             \mp (\tilde{H}_{dL} \pm \tilde{H}_{dR})]/\sqrt{2} \,,
\end{eqnarray}
with mass eigenvalues $m_{\tilde{B}_{1,2}} = M^D_1$, $m_{\tilde{W}^3_{1,2}}=
M^D_2$ and $m_{\tilde{H}_{1,2}}= |\mu|$, respectively. [The coefficient
$\{i\}$ is associated with the second entry in each row.]

The {\underline{neutral}} Majorana fields can be joined pairwise to form three
Dirac fields in the $v= 0$ limit:
\begin{eqnarray}
\tilde\chi^0_D  =
   \{\tilde{B}_D, \tilde{W}^3_D, \tilde{H}_D\}
\end{eqnarray}
where the Dirac fields are expressed in terms of the mass eigenfields as
\begin{eqnarray} \label{mass1}
&& \tilde{W}^3_D = ( \tilde{W}^3_1-i\tilde{W}^3_2)/\sqrt{2}
 \;\;\ \  {\rm and} \;\;\ \    
{\tilde{W}}^3 \Rightarrow {\tilde{B}}     \,, \nonumber\\
 && \tilde{H}^0_D = ( \tilde{H}_1-i\tilde{H}_2)/\sqrt{2}   \,,
\end{eqnarray}
while the corresponding charge-conjugated fields read:
\begin{eqnarray}
&& \tilde{W}^{3c}_D = -( \tilde{W}^3_1+i\tilde{W}^3_2)/\sqrt{2}
 \;\;\ \ {\rm and} \;\;\ \ {\tilde{W}}^{3c} \Rightarrow {\tilde{B}}^c
\,,      \nonumber \\
 && \tilde{H}^{0c}_D = +( \tilde{H}_1 + i\tilde{H}_2)/\sqrt{2}\,.
\end{eqnarray}

The {\underline{charged}} Dirac fields, in parallel to the neutral fields but
in circular notation, are given by
\begin{eqnarray}
&& \tilde{W}^\pm_{1,2}
   = \{i\} [(\tilde{W}'^\pm_L \pm \tilde{W}'^\pm_R)
     \pm (\tilde{W}^\pm_L \pm \tilde{W}^\pm_R)]/\sqrt{2} \,,
     \nonumber\\
&& \tilde{H}^\pm = \tilde{H}^\pm_{uL/R} + \tilde{H}^\pm_{dR/L} \,.
\end{eqnarray}
These fields are mutually conjugate to each other. The $\pm$ fields can be
rotated to three new charged Dirac fields:
\begin{eqnarray} \label{mass2}
{\tilde{\chi}}^\pm_{D1} &=& ({\tilde{W}}^\pm_1 \pm i {\tilde{W}}^\pm_2) /
   \sqrt{2} \,,   \nonumber \\
{\tilde{\chi}}^\pm_{D2}  &=& ({\tilde{W}}^\pm_1 \mp i {\tilde{W}}^\pm_2) /
   \sqrt{2} \,,  \nonumber \\ 
{\tilde{\chi}}^\pm_{D3}  &=& {\tilde{H}}^\pm \,,
\end{eqnarray}
generating, in association of the charged gaugino and higgsino fields, an
ensemble of three chargino fields. Again the $\pm$ components are related by
$C$-conjugation.

In the limit of small but non-zero $v$, all the fields are weakly mixed after
electroweak symmetry breaking, i.e. the original mass eigenfields defined in
Eqs. (\ref{mass1}) and (\ref{mass2}) receive small admixtures. The final
neutralino mass eigenfields may be written, up to terms linear in $v/M_{\rm
  SUSY}$:
\begin{equation}
   \tilde\chi_{\rm phys} \approx
       \begin{pmatrix}
        \mathbb{1}_{4\times 4} & -\Omega_D \\
         \Omega^\dagger_D         & \mathbb{1}_{2\times 2}
        \end{pmatrix} \tilde\chi_{\rm mass}
\end{equation}
with the $4\times 2$ matrix $\Omega_D$ accounting for the admixture between
gauginos and higgsinos,
\begin{equation}
\Omega_D = m_Z \left(\begin{array}{rr}
 \phantom{-} i s_W s_\beta / \mu_{1+}   &  \;  
    \phantom{-i} s_W c_\beta / \mu_{1-}  \\
 - \phantom{i} s_W c_\beta / \mu_{1-}   &  \; -i s_W s_\beta / \mu_{1+} \\
-i c_W s_\beta / \mu_{2+}              &  \; 
- \phantom{i} c_W c_\beta / \mu_{2-}  \\
 \phantom{-i}  c_W c_\beta / \mu_{2-}   &  
\; \phantom{-} i c_W s_\beta / \mu_{2+}
        \end{array}\right)
\label{Dirac_neutralino_omega}
\end{equation}
with $\mu_{1+} = \mu + M^D_1$ etc. For the chargino states, one finds
similarly:
\begin{eqnarray}
   \tilde{\chi}^\pm_{\rm phys} \approx
       \begin{pmatrix}
        \mathbb{1}_{2\times 2} & -\Omega_\pm \\
         \Omega^\dagger_\pm          & 1
        \end{pmatrix} \tilde{\chi}^\pm_{\rm mass}
\end{eqnarray}
with the $2\times 1$ matrix $\Omega_\pm$ taking into account the small mixing
between gauginos and higgsinos,
\begin{eqnarray}
 \Omega_{\pm} = m_W
     \left(\begin{array}{c}
    \phantom{-i} c_\beta / \mu_{2-} \mp  \phantom{i} s_\beta / \mu_{2+}  \\
               -i s_\beta / \mu_{2+} \mp i c_\beta / \mu_{2-}
                 \end{array}\right)\,.
\end{eqnarray}
Up to linear accuracy in $v/M_{\rm SUSY}$ the neutralino and chargino mass
eigenvalues are unaltered. 

The Dirac charge has been introduced for
convenient book-keeping of allowed and forbidden reactions in the N=2 hybrid
theory.  Of course, the charginos form Dirac fields even in the MSSM. However,
for non-zero masses one cannot define a conserved Dirac charge 
in this more restricted theory. The
gauge-strength Yukawa-type couplings of the charginos to a sfermion and an
outgoing left-handed matter fermion involve both the L and R components of the
(current) Dirac wino spinor \cite{Drees}, ${\cal L}^{\rm MSSM}_{\tilde{W}
  \tilde q q} \sim \overline{u_L} \tilde{W}^+_R \tilde d_L + \overline{d_L}
\left( \tilde{W}_L^+ \right)^c \tilde u_L$. In the MSSM these components,
which carry opposite $D-$charge, are coupled by the mass $M_2$. In contrast,
in the N=1/N=2 hybrid theory a conserved Dirac charge (\ref{dc}) can be
defined for $v \to0$, since the L and R components of the original N=1 wino
$\tilde{W}$ belong to {\em different} Dirac fields in this limit. Since no
second ``partner'' field has been introduced in the higgsino sector, their
couplings to fermions and sfermions, which are determined by the standard
Yukawa interactions, will not conserve $D$ either. The transition to mixed
gaugino/higgsino states will be discussed in a sequel to this report.

The formalism can now be applied to compare signatures distinguishing the
original N=1 Majorana theory from a Dirac theory in the electroweak sector
as formulated explicitly in the hybrid model.

\subsection{Electroweak Squark Cascade Decays in Majorana and Dirac Scenarios}
\label{sc:neucasc}

\noindent
The generic structures of sfermion decays to neutralinos/charginos and of
neutralino/chargino decays to sfermion plus fermion pairs are similar to those
of squark and gluino decays in super- and hyper-QCD. The complexity increases
due to the mixing between gauginos and higgsinos and between left- and
right-handed sfermions originating from electroweak symmetry breaking. However,
for the first and second generation (s)fermions with small Yukawa couplings the
contamination is negligible.

A rich ensemble of observables for measuring the properties of supersymmetric
particles at the LHC is provided by cascade decays involving neutralinos. In
particular, the squark cascades with intermediate neutralinos and sleptons
have served to study experimental prospects of measuring masses and spins.  In
addition, the Majorana or Dirac nature of the neutralinos can be determined by
measuring the distributions of the charged leptons in the final state.

In the following discussion we assume that only SU(2) singlet sleptons
$\tilde l_R$ are accessible in the decay of the relevant neutralino $\tilde
\chi_2^0$.  Ignoring lepton mass effects, the charged ``near'' lepton produced
together with the slepton is then either a left-handed $l^+$ or a
right-handed $l^-$. [We will see in a moment that only one of these
possibilities is allowed in the Dirac theory.] Neutralinos produced in $\tilde
q_L$ decays are produced in association with a left-handed quark, i.e. they
are predominantly left-handed. Angular momentum conservation then implies
that a near $l^-$ [$l^+$] preferentially goes opposite [parallel] to the
neutralino flight direction. In the rest frame of the decaying $\tilde q_L$ a
near $l^-$ will thus tend to be softer, and closer to the quark in phase
space, than a near $l^+$.  These correlations are reflected in the
invariant $q l$ mass distributions \cite{Barr:2004ze,lhcspin}. The same
argument implies that the slepton, and hence the ``far'' lepton that results
from its decay, will be harder [softer] if it has positive [negative] charge.

In {\underline{Majorana}} theories the neutralino $\tilde\chi^0_2$ can decay
into sleptons $\tilde l_R$ of both positive and negative charge:
\begin{equation}
   \tilde{q}_L \to q\, \tilde\chi^0_2 \to q\, l^\mp_n \tilde{l}_R^\pm
               \to q\, l^\mp_n \, l^\pm_f \, \tilde\chi^0_1 \,.
\label{eq:chain1}
\end{equation}
The near $(n)$ leptons and the far $(f)$ leptons, produced directly in the
$\tilde\chi^0_2$ decays and in the subsequent $\tilde l_R$ decays
respectively, both can have either negative or positive charges, albeit with
different energy distributions as a result of the neutralino polarization
discussed above.

By contrast, the transition from Majorana to Dirac particles leads to a
simpler situation. In the {\underline{Dirac}} theory, evaluating the generic
fermion-sfermion-neutralino Lagrangian, restricted to gauginos for the first
two generations, results in 
\begin{equation}
{\mathcal{L}}^{f \tilde{f} \tilde{\chi}} =
      g_L \left(\overline{f_L} \, \tilde\chi^0_D \, \tilde{f}_L
          + {\overline{{\tilde\chi}^0_{D}}} \, f_L\, \tilde{f}_L^\ast \right)
       + g_R \left(\overline{f_R}\, \tilde\chi_D^{c0} \, \tilde{f}_R
              + {\overline{{\tilde\chi}_{D}^{c0}}} \,
              f_R \, \tilde{f}_R^\ast\right)\,.
\end{equation}
[The L- and R-couplings $g_L, g_R$ are defined in terms of the neutralino
mixing matrix and the fermion isospin and hyper-charges, as frequently noted
in the literature; recall that $D[{\tilde{\chi}}^0_D] = -
D[{\tilde{\chi}}^{c0}_D] = +1$.] A fixed sequence of charges in leptonic decay
modes is thus predicted. The squark decay generates, together with the quark,
an antineutralino ${\tilde\chi}^{c0}_{D2}$, the antineutralino in turn
decays to a lepton $l^-$ and an antislepton ${\tilde l}^+_R$ which
finally decays into an antilepton $l^+$:
\begin{equation}
   \tilde{q}_L \to q\, {\tilde\chi}^{c0}_{D2}
               \to q\, l^-_n\, \tilde l^+_R
               \to q\, l^-_n\, l^+_f\, {\tilde\chi}^{c0}_{D1}  \,.
\label{eq:chain2}
\end{equation}
In other words, only one of the two possibilities available in the Majorana
theory can be realized in the Dirac theory.

\begin{figure}
\flushleft
\fbox{\makebox[17.5cm]{\scriptsize --------
$\;\; \tilde{q}_L \to q \tilde\chi^0_2 \to q  l^\mp_n  \tilde l_R^\pm
               \to q l^\mp_n l^\pm_f  \tilde\chi^0_1$
\hfill
-- -- --
$\;\;\; \tilde{q}_L \to q \tilde\chi^{c0}_{D2}
               \to q l^-_n \tilde l^+_R
               \to q l^-_n l^+_f \tilde\chi^{c0}_{D1}$
\hfill
$\cdot\cdot\cdot\cdot\cdot$
$\;\; \tilde{q}_L^* \to \bar{q} \tilde\chi^0_{D2}
               \to \bar{q} l^+_n \tilde l^-_R
               \to \bar{q} l^+_n l^-_f \tilde\chi^0_{D1}$}}%
\\[3ex]
\epsfig{figure=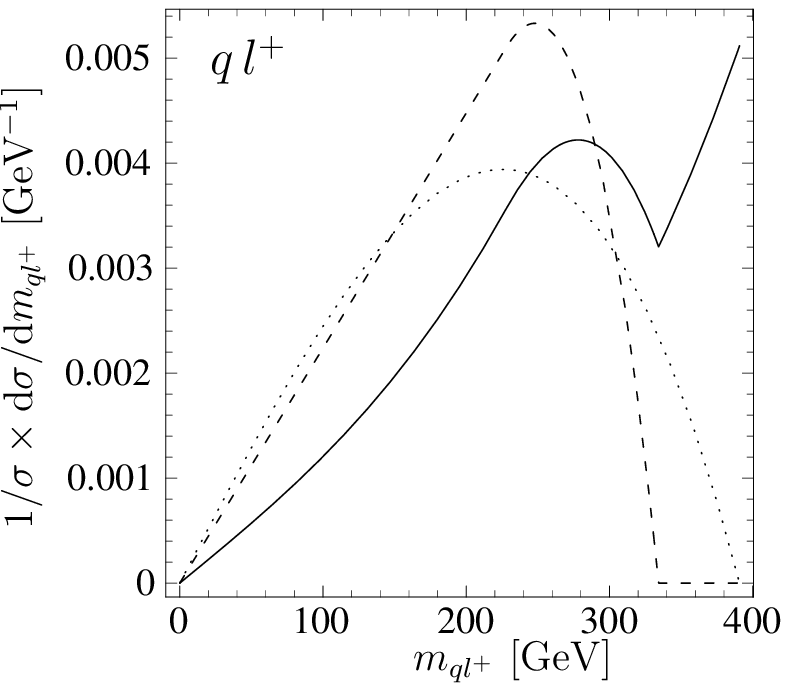, width=5.7cm}
\hfill
\epsfig{figure=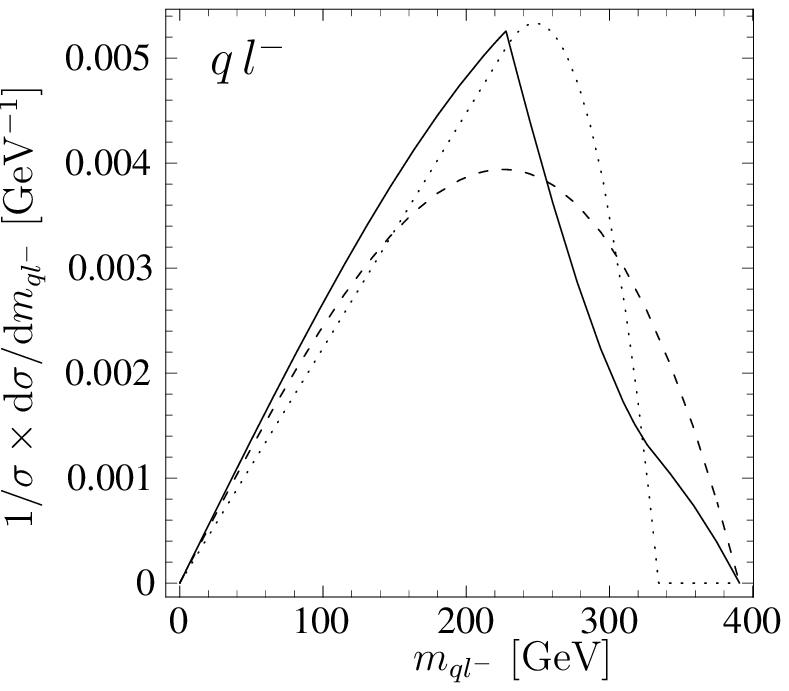, width=5.7cm}
\hfill
\epsfig{figure=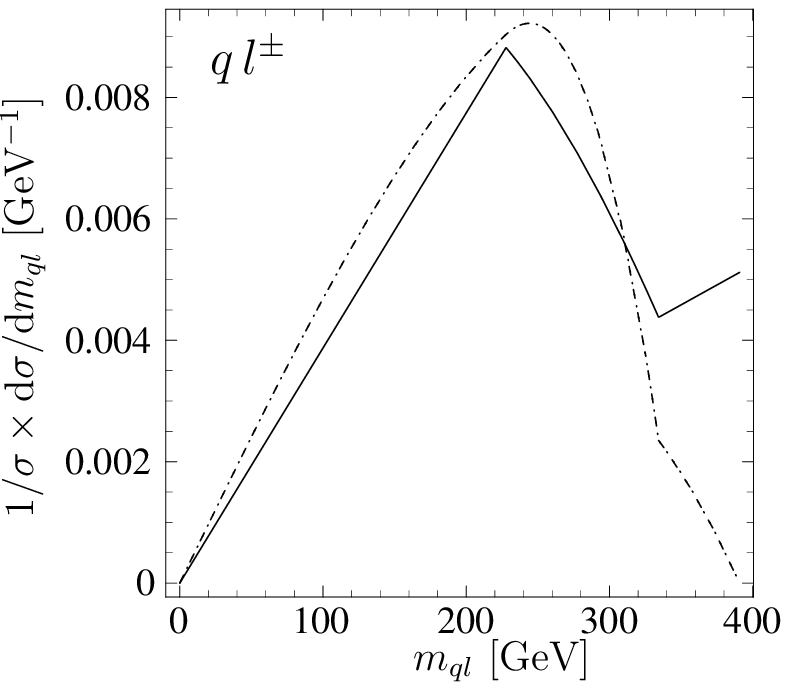, width=5.7cm}
\vspace{-2ex}
\caption{$ql$ invariant mass distributions for squark decay chains involving
Majorana or Dirac neutralinos. The masses have been taken from the SPS1a$'$
scenario \cite{SPS1a}.}
\label{fig:ql}
\end{figure}
Following the calculations of Ref.~\cite{lhcspin} we derive the $q l^-$ and
$q l^+$ distributions as shown in Fig.~\ref{fig:ql} for the decay chains in
Eqs.~\eqref{eq:chain1} and \eqref{eq:chain2}. In order to understand these
figures, note first that for the given choice of superparticle masses the
endpoint of the $q l_n$ invariant mass distribution is larger than that for
the $q l_f$ distribution. Comparison of the solid curves in the first two
frames clearly shows that, for $\tilde q_L$ decay, the $q l^+$ distribution
is significantly harder than the $q l^-$ distribution; recall that this is
true for both the near and far lepton. Turning to the Dirac scenario, we saw
that the $l^+$ from $\tilde q_L$ decay has to be the far lepton; the dashed
curve in the first frame therefore cuts off at the lower $q l_f$ endpoint.
Note that this distribution is indeed quite hard, i.e. it peaks fairly close
to this endpoint. In contrast, the dashed curve in the second frame shows the
distribution of the near lepton in $\tilde q_L$ decay. Since this lepton is
negatively charged, the above discussion leads us to expect this distribution
to be relatively soft, and indeed it peaks well below its endpoint.

As demonstrated in the figure, the invariant mass distributions are markedly
different for the Dirac cascade scenario compared to the Majorana cascade
scenario. Though the sensitivity is reduced to some extent, this is true even
when the charge of the lepton is undetermined, as a result of the polarization
of the $\tilde\chi^0_2$ stemming from the squark decay.  Quite generally, the
charge conjugated process
\begin{equation}
   \tilde{q}_L^* \to \bar{q}\, \tilde\chi^0_{D2}
               \to \bar{q}\, l^+_n\, \tilde l^-_R
               \to \bar{q}\, l^+_n\, l^-_f\, \tilde\chi^0_{D1}
\label{eq:chain3}
\end{equation}
leads, by $CP$-invariance, to charge-chirality correlations exactly opposite
to Eq.~\eqref{eq:chain2}, so that the $q l^\pm$ distribution from
$\tilde{q}_L$ decays is identical to the ${\bar{q}} l^\mp$ distribution
from $\tilde{q}_L^*$ decays.  As a result, the $q l^\pm$ spectrum, in
contrast to the $q l^+$ and $q l^-$ spectra, is insensitive to the
squark charge so that the analysis of this distribution, not requiring
knowledge of the parton distribution functions, is particularly
simple.\footnote{At the LHC one expects more $\tilde q_L$ than $\tilde q_L^*$
  to be produced, i.e. the charge averaging should be done with different
  weights. This would increase the difference between the two theories even
  further. Also note that these distributions can be measured directly
  only for $l=e,\mu$.}

\subsection{Electroweak Majorana signatures in \boldmath $e^-e^-$ collisions}

\noindent
Polarized electron-electron collisions \cite{Keung:1983nq,gudi} offer a
classical and most transparent method for studying the Majorana character of
neutralinos:
\begin{eqnarray}
   e^-_L e^-_L &\to& \tilde{e}^-_L \tilde{e}^-_L \,, \;\;
                     e^-_R e^-_R \to \tilde{e}^-_R \tilde{e}^-_R  \nonumber  \\
   e^-_L e^-_R &\to& \tilde{e}^-_L \tilde{e}^-_R                            \,.
\end{eqnarray}
All three processes are activated in Majorana theories while, in analogy to
$qq$ scattering, the equal-helicity amplitudes vanish for Dirac neutralino
exchange. Electron beams can be polarized at linear colliders to nearly 100\%
and, as a minor idealization, we will assume complete polarization for the
sake of clarity in the following analysis [corrections to this assumption can
trivially be implemented].

In the hybrid theory on which we have based the detailed analyses, the
scattering amplitudes can be written as:
\begin{eqnarray}
A[e^-_L e^-_L \to \tilde{e}^-_L \tilde{e}^-_L]
 &=& - 2 e^2 \left[\mathcal{M}_{LL}(s,t) + \mathcal{M}_{LL}(s,u)\right]\,,
     \nonumber\\
A[e^-_R e^-_R \to \tilde{e}^-_R \tilde{e}^-_R]
 &=&  2 e^2 \left[\mathcal{M}^*_{RR}(s,t) + \mathcal{M}^*_{RR}(s,u)\right]\,,
     \nonumber\\
A[e^-_L e^-_R \to \tilde{e}^-_L \tilde{e}^-_R]
 &=& e^2 \lambda^{1/2} \sin\theta \; \mathcal{D}_{LR}(s,t)\,.
\end{eqnarray}
Here $\theta$ is the scattering angle, and the dimensionless neutralino
functions $\mathcal{M}_{ab}$ and $\mathcal{D}_{ab}$ ($a,b=L,R$) are defined by
\begin{eqnarray}
\mathcal{M}_{ab}(s,t(u))
   &=& \sum^6_{k=1} \frac{m_{\tilde{\chi}^0_k}}{\sqrt{s}}
       \mathcal{V}_{ak} \mathcal{V}_{bk} D_{kt(u)}\,,\nonumber\\
\mathcal{D}_{ab}(s,t(u))
   &=& \sum^6_{k=1} \mathcal{V}_{ak} \mathcal{V}^*_{bk} D_{kt(u)}\,,
\end{eqnarray}
They are determined by the $t(u)$-channel neutralino propagators $D_{kt(u)} =
s/(t(u)- m^2_{\tilde{\chi}^0_k})$ and the effective mixing coefficients
\begin{eqnarray}
\mathcal{V}_{Lk} &=& \mathcal{N}_{2k}/2c_W
                  +\mathcal{N}_{4k}/2 s_W\,,\nonumber\\
\mathcal{V}_{Rk} &=& \mathcal{N}_{2k}/c_W\,.
\end{eqnarray}
The neutralino mixing matrix $\mathcal{N}$ diagonalizes the neutralino mass
matrix as $\mathcal{N}^T \mathcal{M}_n\, \mathcal{N} ={\rm
  diag}(m_{\tilde{\chi}^0_1},\ldots, m_{\tilde{\chi}^0_6})$. The differential
cross sections,
\begin{eqnarray}
\frac{{\rm d}\sigma_{LL}}{{\rm d}\cos\theta}
 &=&
\frac{\pi\alpha^2}{4s}
\lambda^{1/2}\, \left|\mathcal{M}_{LL}(s,t)+\mathcal{M}_{LL}(s,u)\right|^2\,,
      \nonumber \\
\frac{{\rm d}\sigma_{RR}}{{\rm d}\cos\theta}
 &=&
\frac{\pi\alpha^2}{4s}
\lambda^{1/2}\, \left|\mathcal{M}_{RR}(s,t)+\mathcal{M}_{RR}(s,u)\right|^2\,,
      \nonumber \\
\frac{{\rm d}\sigma_{LR}}{{\rm d}\cos\theta}
 &=&
\frac{\pi\alpha^2}{4s}
\lambda^{3/2}\, \sin^2\theta
\left|\mathcal{D}_{LR}(s,t)+\mathcal{D}_{LR}(s,u)\right|^2\,,
\label{ee_to_sese}
\end{eqnarray}
can easily be derived from the scattering amplitudes.

In the standard Majorana limit the expressions reduce to the familiar MSSM
form, see e.g. Ref.~\cite{slep}. The differential cross sections are the same
in their form as those in Eq.~\eqref{ee_to_sese}, with the $t$- and
$u$-channel exchanges mediated only by the four mass eigenstates
$\tilde{\chi}^0_{1,3,5,6}$; the other two states $\tilde{\chi}_{2,4}$ are
decoupled as $M'_{1,2}$ become infinite.

\begin{figure}
\epsfig{figure=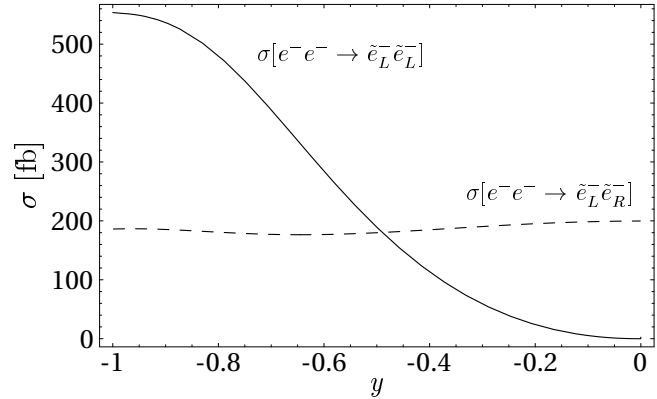, width= 8.5cm}
\vspace{-2ex}
\caption{Partonic cross sections for same-sign selectron production as a
functions of the Dirac/Majorana control parameter~$y$, for $\sqrt{s} = 500$ GeV
and SPS1a$'$ parameters \cite{SPS1a}.
Not shown is the cross section for $e^-e^- \to \tilde{e}^-_R \tilde{e}^-_R$,
which, apart from the different normalization, shows a similar behavior as the
cross section for $e^-e^- \to \tilde{e}^-_L \tilde{e}^-_L$.}
\label{fig:sese}
\end{figure}

The Dirac limit, on the other hand, is exceptionally simple in the selectron
sector in which the Yukawa couplings $\sim m_e / v$ can be neglected.  The
higgsino couplings vanish in this limit and the higgsino admixtures to the
U(1) and SU(2) gauginos are ineffective.  Hence the neutralino system is
isomorphic, apart from the SU(3) symmetry group, to the gluino system. The
differential cross sections in the Dirac limit with the gaugino and higgsino
mixing neglected greatly simplify to
\begin{eqnarray}
\frac{{\rm d}\sigma_{LL}}{{\rm d}\cos\theta}
  &=&\frac{{\rm d}\sigma_{RR}}{{\rm d}\cos\theta}= 0\,,  \nonumber \\
\frac{{\rm d}\sigma_{LR}}{{\rm d}\cos\theta}
  &=& \frac{{\rm d}\sigma_{RL}}{{\rm d}\cos\theta}
   = \frac{\pi\alpha^2}{16 c^4_W s} \lambda^{3/2} \sin^2\theta
     \left(\frac{s}{t-m^2_{\tilde{\chi}^0_1}}
         +\frac{s}{u-m^2_{\tilde{\chi}^0_1}}\right)^2 \,.
\end{eqnarray}
The two representative cross sections $\sigma_{LL}$ and $\sigma_{LR}$ are
shown along the path $\mathcal{P}$, defined analogously to the QCD sector, in
Fig.~\ref{fig:sese}. In the figure, gaugino and higgsino mixing induced by
electroweak symmetry breaking has been included by diagonalizing the complete
mass matrix \eqref{eq:neu_mass_matrix} numerically, but the quantitative
effect of this mixing is very small.

In Ref.~\cite{AguilarSaavedra:2003hw} a detailed phenomenological analysis for
selectron production in $e^-e^-$ collisions was performed. It was shown that,
by using different decay modes of the selectrons, their masses can be
reconstructed experimentally, thus allowing a clear distinction between the
processes $e^-e^- \to \tilde{e}^-_R \tilde{e}^-_R, \; \tilde{e}^-_R
\tilde{e}^-_L, \; \tilde{e}^-_L \tilde{e}^-_L$. Therefore the Majorana nature
of the neutralinos with dominant gaugino component can be tested unambiguously
in $e^-e^-$ collisions.

\section{Like-Sign Dileptons and Unlike-Sign Dileptons at the LHC}
\label{sec:lhc_phenomenology}

\noindent
In the previous sections, two methods have been identified for the
experimental discrimination between Dirac and Majorana gauginos at the LHC:
The correlation between charge and helicity of fermions from Dirac neutralino
decays leaves a characteristic imprint on the quark-lepton distributions, as
shown in Section~\ref{sc:neucasc}, which cannot be the result of modifications
in the sparticle spectrum.  Secondly, the production cross sections for
squarks and gluinos are different in the two cases, as analyzed in
Section~\ref{sc:hyperqcd}. In the following it will be shown how this
difference can be measured through like-sign and unlike-sign dilepton signals
at the LHC. Before describing the detailed phenomenological analysis for the
rates of like-sign dilepton events, a general overview of like-sign and
unlike-sign dileptons will be given to set the frame for expectations in
various channels of the sub-processes.

\subsection{A Coarse Picture of Like-Sign and Unlike-Sign Dilepton Channels}

\begin{figure}
\psfig{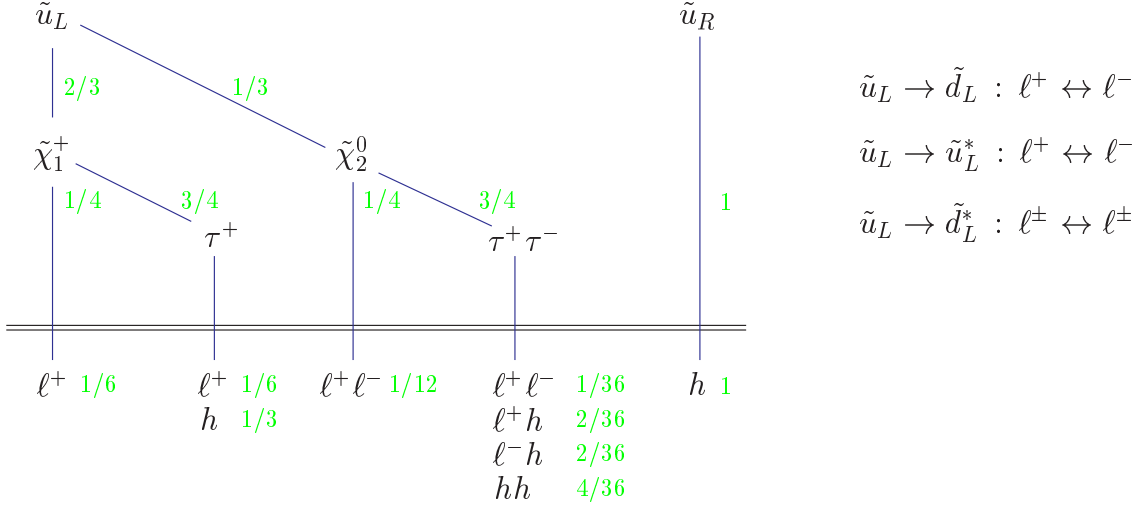}
\caption{Sketch of allowed decay chains for L-type
  and R-type $u$ squarks, $\tilde{u}_{L,R}$, for SPS1a$'$ masses. Here $h$
  stands for fully hadronic decay channels without charged leptons, while
  $\ell$ stands for an electron or muon. The numbers in light green/gray
  denote approximate branching ratios for the associated decay channels. The
  decay patterns for the other (anti)squarks can be derived by the
  replacements given on the right side. } 
\label{fig:decay_chain}
\end{figure}

\noindent
To get a transparent view of channels which allow us to confront the Majorana
nature of the gluinos with the Dirac alternative, we will first consider
characteristic examples, focusing on the ratio of like-sign dilepton events of
different charge and the ratio of like-sign over unlike-sign dileptons.
Like-sign lepton pairs can be produced from decays of L-squark pairs mediated
by charginos, e.g. for $\tilde{u}$ and $\tilde{d}$-squarks
\begin{eqnarray}
\tilde{u}_L &\to& d \, \tilde{\chi}^+_1 \to d \, l^+ \nu_l \, 
\tilde{\chi}_1^0         \nonumber \\
\tilde{d}_L &\to& u \, \tilde{\chi}^-_1 \to u \, l^- {\bar{\nu}}_l \, 
\tilde{\chi}_1^0 \,,
\label{eq:chain}
\end{eqnarray}
as sketched in Fig.$\,$\ref{fig:decay_chain}. For easy lepton and charge
identification, we restrict ourselves to $l=e,\mu \equiv \ell$, or $l=\tau$
with leptonic tau decays $\tau \to e \nu \bar{\nu},\; \mu\nu \bar{\nu}$.
Owing to the valence quark distribution in the proton beams, $\ell^+\ell^+$
and $\ell^-\ell^-$ pairs are not produced in equal numbers in SUSY events.
Decay chains with neutralinos, on the other hand,
\begin{equation}
\tilde{u}_L \to u \, \tilde{\chi}^0_2 \to u \, l^+ l^- \, 
\tilde{\chi}_1^0 \,,
\end{equation}
lead to predominantly opposite-sign and same-flavor leptons in the final
state. [They give only a small contamination to the like-sign dilepton signal
when mixed lepton-hadron decays of neutralinos to tau pairs are observed, or,
for experimental reasons, when one of the leptons is missed in the detector.]
An overview of like-sign and unlike-sign dilepton ratios is presented in
Tab.$\,$\ref{tab:majorana_dirac_overview}.

For specifying the decay branching ratios, the reference scenario
SPS1a$'$~\cite{SPS1a} will be adopted. In this scenario, BR$[\tilde{q}_L \to
q' \, \tilde{\chi}^\pm_1] \approx 2/3$, BR$[\tilde{q}_L \to q \,
\tilde{\chi}^0_2] \approx 1/3$ and BR$[\tilde{q}_R \to q \, \tilde{\chi}^0_1]
\approx 1$, which is typical for scenarios with wino-like $\tilde{\chi}^0_2$
and bino-like $\tilde{\chi}^0_1$.  The charginos ${\tilde{\chi}}^\pm _1$ and
the neutralino ${\tilde{\chi}}^0 _2$ decay preferentially to taus with
branching ratios $\approx 3/4$.

\begin{table}[htb]
\renewcommand{\arraystretch}{1.35}
\begin{tabular}{|c||r|r|r||r|r|r|}
\hline
Process & \multicolumn{3}{c||}{Majorana}
        & \multicolumn{3}{c|}{Dirac} \\
   \cline{2-7}
        & $\ell^+\ell^+$
        & $\ell^-\ell^-$
        & $\ell^+\ell^-$
        & $\ell^+\ell^+$
        & $\ell^-\ell^-$
        & $\ell^+\ell^-$ \\
\hline\hline
$u_L u_L \to \tilde{u}_L \tilde{u}_L$ &
       49 &  1 & 46 & $\times$  & $\times$  & $\times$  \\
$d_L d_L \to \tilde{d}_L \tilde{d}_L$ &
       1 &  49 & 46 & $\times$  & $\times$  & $\times$  \\
$u_L d_L \to \tilde{u}_L \tilde{d}_L$ &
       7 &  7 & 82 & $\times$  & $\times$  & $\times$   \\ \hline
$u_L q_R \to \tilde{u}_L \tilde{q}_R$ &
        0 & 0 & 36 & 0 & 0 & 36  \\
                                                          \hline
$q_L \bar{q}_L \to \tilde{q}_L \tilde{q}^*_L$ &
      7 & 7  & 82  & 7 & 7 & 82 \\
$u_L \bar{d}_L \to \tilde{u}_L \tilde{d}^*_L$  &
      49 &  1 & 46 & 49  & 1  & 46 \\
$u_L \bar{q}_R \to \tilde{u}_L \tilde{q}^*_R$ &
        0 & 0 & 36 & $\times$ & $\times$ & $\times$  \\
                                                                \hline
$g u_L  \to \tilde{g}_{(D)} \tilde{u}_L $ &
       14 & 2 & 50 & 14 & 2 & 50 \\
$g \bar{u}_L  \to \tilde{g}_{(D)}^{(c)} \tilde{u}^*_L $ &
       2 & 14 & 50 & 2 & 14 & 50 \\
$g q_R  \to \tilde{g}_{(D)}^{(c)} \tilde{q}_R $ &
      0 & 0 & 18 & 0 & 0 & 18  \\
$g \bar{q}_R  \to \tilde{g}_{(D)} \tilde{q}^*_R $ &
      0 & 0 & 18 & 0 & 0 & 18  \\ \hline
$g g \to \tilde{g}_{(D)} \tilde{g}_{(D)}^{(c)}$ &
      4  & 4  &  34 &  4 & 4 & 34 \\
\hline
\end{tabular}
\caption{Approximate relative probabilities of like-sign lepton pairs $\ell^+\ell^+$
      and $\ell^-\ell^-$, and unlike-sign lepton pairs $\ell^+\ell^-$,
         separately for characteristic channels [$q=u$ or $d$]; the proper
         normalization of the probabilities requires dividing all entries by
         the common denominator ${\mathcal{N}}_n = 324$. 
         Probabilities for $\tilde{d}$ processes which can be derived by
         isospin rotation of $\tilde{u}$ processes are not noted explicitly. 
         Parton processes forbidden in the Dirac theory are marked by the
         symbol $\times$.} 
\label{tab:majorana_dirac_overview}
\end{table}

\paragraph*{(a) \underline{Squark pair production:}}\

In the Majorana theory the most prominent squark production channels are the
subprocesses $u u \to \tilde{u} \tilde{u}$, $d d \to \tilde{d} \tilde{d}$ and
$u d \to \tilde{u} \tilde{d}$, initiated by valence quarks and mediated by
gluino exchange. In the Majorana theory, the $\tilde{u}_L \tilde{u}_L$ and
$\tilde{d}_L \tilde{d}_L$ pair production processes lead to same-sign leptons,
whereas opposite-sign dileptons are generated in $\tilde{u}_L \tilde{d}_L$
events, if both squarks decay into charginos. In both the Dirac and Majorana
theory opposite-sign dileptons can originate from ${\tilde{q}}_L
{\tilde{q}}_R$ final states via ${\tilde{\chi}}^0_2 \to \ell^+ \ell^-$ and
hadronic decays of ${\tilde{q}}_L$ and ${\tilde{q}}_R$ squarks, respectively.
The following event fractions and ratios
\begin{align}
N(\ell^+\ell^+) / N(\ell^-\ell^- )     & \sim  3     \qquad \mbox{(Majorana)}
   \nonumber \\
N(\ell^\pm\ell^\pm) / N(\ell^+\ell^-)  & \sim  1/4
\intertext{and}
N(\ell^\pm\ell^\pm) / N(\ell^+\ell^-)  & = 0 \qquad \mbox{(Dirac)}
\end{align}
are obtained for $(2u+d)$ valence partons in the proton.

In both the Majorana and the Dirac theory, squark pairs can also be produced
from quark-antiquark scattering and gluon annihilation. The dominant
contributions for dileptons come from the processes, $u_L \overline{u_L} \to
\tilde{u}_L\tilde{u}^*_L$, $d_L \overline{d_L} \to \tilde{d}_L \tilde{d}^*_L$,
$u_L \overline{d_L} \to \tilde{u}_L \tilde{d}_L^*$ and $d_L \overline{u_L} \to
\tilde{d}_L \tilde{u}_L^*$ .
These channels have one valence quark and one sea antiquark in the initial
state, so that the cross sections are smaller than the quark-quark cross
sections. The channels predict a ratio $N(\ell^+\ell^+)/N(\ell^-\ell^-) \sim
2$, for an approximate fraction 1/4 of like-sign events within the total
dilepton sample.  These leading channels are not altered by switching from the
Majorana to the Dirac theory.

Channels that are initiated by two sea (anti)quarks are doubly suppressed.

\paragraph*{(b) \underline{Super-Compton Process:}}\

Gluinos $\gw$ decay in the Majorana theory democratically at equal rates to
$\tilde{u}, \tilde{u}^\ast$ and $\tilde{d}, \tilde{d}^\ast$ squarks, both L-
and R-types, of the first two generations. Therefore the super-Compton process
$q g \to \tilde{q} \tilde{g}$ generates like-sign leptons with a branching
ratio that is independent of the squark charge. [Second generation
$\tilde{s},\tilde{c}$ squarks will be included in the subsequent
phenomenological analysis.]  However, since the super-Compton process is
predominantly initiated by valence quarks, positively charged like-sign
leptons pairs outnumber negatively charged like-sign pairs by the ratio
$N(\ell^+\ell^+)/N(\ell^-\ell^-) \sim 2$. The number of unlike-sign dilepton
events is dominant, the ratio $N(\ell^+\ell^+ +\ell^-\ell^-)/N(\ell^+\ell^-)
\sim 1/4$, due to the additional enhancement of the final states generated by
${\tilde{q}}_L \to {\tilde{\chi}}^0_2$ decays accompanied by non-leptonic
${\tilde{q}}_R$ jet decays.

The picture becomes a bit more subtle when switching to the Dirac theory.
L-squarks $\tilde{q}_L$ are only produced together with $\gw_D$ gluinos,
whereas $\sigma[gq \to \qw_L \gw^c_D] = 0$, see Eq.~\eqref{eq:xsec_comp2}.
According to Eqs.~\eqref{eq:gldec1} and \eqref{eq:gldec2}, the $\gw_D$ gluinos
subsequently decay only into $\tilde{u}^*_L,\tilde{d}^*_L$, but not into
$\tilde{u}_L,\tilde{d}_L$.  Nevertheless, since gluinos decay democratically
to each flavor (anti)-squark for equal masses, the probabilities of the
like-sign and unlike-sign lepton pairs are not altered by switching from the
Majorana to the Dirac theory. [For $\tan\beta > 1$, $\tilde d_L$
  squarks are slightly heavier than $\tilde u_L$ squarks, so that the latter
  are slightly preferred in gluino decays. This small effect has been taken
  into account in the numerical analysis to be described in the following
  subsection.]

\paragraph*{(c) \underline{Gluino Pair-Production:}}\

Pair production of gluinos in the Majorana theory leads to same-sign L-squark
pairs ($\tilde{u}_L \tilde{u}_L$, $\tilde{u}^*_L \tilde{u}^*_L$, $\tilde{d}_L
\tilde{d}_L$, $\tilde{d}^*_L \tilde{d}^*_L$, $\tilde{u}_L \tilde{d}_L^*$ and
$\tilde{u}^*_L \tilde{d}_L$) in half of the cases, which in turn generate
same-sign leptons pairs through the chargino decay chain, with the charge
ratio $N(\ell^+\ell^+)/N(\ell^-\ell^-) \sim 1$. In the Dirac theory, only
$\gw_D \gw^c_D$ gluino pairs are generated and $\gw_D$ decays only into
L-antisquarks $\tilde{q}_L^*$ (and R-squarks $\tilde{q}_R$) while $\gw_D^c$
decays only into L-squarks $\tilde{q}_L$ (and R-antisquarks
$\tilde{q}_R^\ast$).  However, as for case {\it (b)}, owing to the
flavor-democratic decays of (Dirac) gluinos the relative rates for like-sign
and unlike-sign lepton pairs are unchanged for the Dirac theory compared to
the Majorana theory. In other words, contrary to popular belief the frequent
occurrence of like-sign dilepton pairs in gluino pair events is {\em not} a
signal for the Majorana nature of the gluino.

The total production cross section for gluino pairs is roughly twice as large
in the Dirac theory compared to the Majorana theory, as a result of the
doubling of the physical degrees of freedom of the gluinos.

\paragraph*{\underline{In summary},} the population of like-sign dileptons
predicted in the Majorana theory is altered significantly when switching to
the Dirac theory, with the suppression of like-sign as well as unlike-sign
dileptons in the valence channels being most prominent. Properly weighing the
individual channels,
\begin{equation}
   \ell^+\ell^+ / \ell^- \ell^- / \ell^+\ell^-  =
   \sum \sigma_k f_k^{\pm\pm/+-} / \sum \sigma_k    \,,
\end{equation}
the valence-valence and super-Compton channels generate the leading
contributions, of similar size as demonstrated in the next subsection.  In
addition to the absolute rates for $\ell^+\ell^+$ and $\ell^-\ell^-$
production, it is very useful to tag the large transverse momentum jets in the
like-sign dilepton events, since this observation allows us to discriminate
between squark and gluino production as the primary hard process
\cite{yuk,yuk2}. Thus, detailed analyses of dilepton events can provide
powerful discriminants between the Majorana and Dirac nature of the gluinos.

\subsection{A Detailed Analysis of Like-Sign Dileptons in Majorana/Dirac
  Theories} 

\noindent
Since our numerical analysis of like-sign dileptons follows strictly the
report on the measurement of the Yukawa coupling in super-QCD, we will not
repeat any of the technical points described comprehensively in
Refs.~\cite{yuk,yuk2}. For representative numerical results, we will adopt the
MSSM scenario of this study, which is close to the reference points
SPS1a$'$~\cite{SPS1a} and the Snowmass point SPS1a~\cite{SPS}. Though the
supersymmetry mass spectrum is comparatively light, it is compatible,
nevertheless, with analyses of high precision electroweak measurements
\cite{ellisea}. Higher supersymmetric masses reduce the production rates and
would thus require larger integrated luminosities at the LHC to obtain similar
event numbers. We note, however, that the processes initiated purely by
valence quarks will drop off most slowly. Heavier spectra thus mean less
``pollution'' of the SUSY dilepton sample by events with gluinos in the final
state; we saw above that ratios of dilepton final states in these gluino
events are identical in the Dirac and Majorana theories. Increasing the
sparticle masses should therefore reduce the number of events needed to
cleanly distinguish between the two theories. 

The masses and branching ratios for squarks, gluinos and charginos/neutralinos
of SPS1a$'$ are tabulated in Ref.~\cite{yuk,yuk2}. In this scenario, gluinos
are heavier than squarks so that they decay via $\gw \to \tilde{q}\,\bar{q},
\, \tilde{q}^*q$.  The branching ratios involving charginos and neutralinos
are not altered when switching from the Majorana to the Dirac theory, except
for the charge-helicity correlations discussed in Section~\ref{sc:neucasc},
which however do not matter for this analysis.\footnote{The simulation has
  been performed using PYTHIA \cite{pythia}, which does not keep track of the
  polarization of decaying neutralinos. Since we use very mild cuts on the
  charged leptons, these polarization effects should not change the event
  numbers significantly.}

Based on the parton cross sections derived in the preceding sections, the
theoretical predictions for the $pp$ cross sections at the LHC are summarized in
Tab.~\ref{tab:xsec}. The values are given, in the Majorana as well as the
Dirac theory, for the relevant squark and gluino channels.  Parallel to
Ref.~\cite{yuk2} a set of cuts has been applied to fight the huge background
cross section from the Standard Model processes: at least two jets with
$p_{\rm T,j} > 200$ GeV, missing transverse energy $\Emiss_{\rm T} > 300$ GeV,
exactly two isolated same-sign leptons $\ell =e,\mu$ with $p_{\rm T,\ell} > 7$
GeV, and a bottom-flavor veto. After applying the cuts, this SM background is
suppressed to a level of 5\%.
\begin{table}
\renewcommand{\arraystretch}{1.35}
\begin{tabular}{|l||r|r||r|r||r|r|}
\hline
\,\, Process & \multicolumn{2}{c||}{Majorana}
 & \multicolumn{2}{c||}{Dirac}
 & \multicolumn{2}{c|}{$N(\ell^+\ell^+)/N(\ell^-\ell^-)$}\\
\cline{2-7}
        & Total cross-section & With BRs and cuts &
          Total cross-section & With BRs and cuts &
               Majorana       &    Dirac          \\
\hline\hline
$\sigma[\tilde{q}_L \tilde{q}_L^{(\prime)}]$ &
    2.1 pb & 6.1 fb & 0 & 0 & 2.5  & $-$  \\
$\sigma[\tilde{q}_L \tilde{q}_L^{(\prime)*}]$ &
    1.4 pb & 3.1 fb & 1.4 pb & 3.1 fb & 1.4 & 1.4 \\
$\sigma[\tilde{q}_L\gw_{(D)}^{\phantom{()}}]$ &
    7.0 pb & 7.6 fb & 7.0 pb & 7.6 fb & 1.5 & 1.5 \\
$\sigma[\gw_{(D)}^{\phantom{()}}\gw_{(D)}^{(c)}]$ &
    3.2 pb & 1.4 fb & 7.0 pb & 3.2 fb & 1.0 & 1.0 \\
\hline
$\sigma$[SM] & 800 pb & $<$0.6 fb & 800 pb & $<$0.6 fb &
\multicolumn{2}{c|}{1.0} \\
\hline
\end{tabular}
\caption{Signal and background cross-sections before and after including
branching ratios (BRs) and applying the
cuts of Ref.~\cite{yuk2}. The numbers always include also the charge conjugate
of the processes in the first column.}
\label{tab:xsec}
\end{table}
Also shown in the table is the ratio of reconstructed positively and
negatively charged lepton pairs, $N(\ell^+\ell^+)/N(\ell^-\ell^-)$, for each
of the production channels. As a result of the more realistic simulation, the
values for $N(\ell^+\ell^+)/N(\ell^-\ell^-)$ are washed out compared to the
naive estimates of the previous subsection.

For an integrated luminosity of $\int \! {\mathcal{L}} = $ 300 fb$^{-1}$ at
the LHC one obtains the following event numbers for the final states with $\ell^+
\ell^+$ and $\ell^- \ell^-$,
\begin{align}
\mbox{Majorana:}& \qquad N(\ell^+\ell^+) = 3,500 && N(\ell^-\ell^-) = 2,100 &&
N(\ell^+\ell^+)/N(\ell^-\ell^-) = 1.66 \\[1ex]
\mbox{Dirac:}& \qquad N(\ell^+\ell^+) = 2,400 && N(\ell^-\ell^-) = 1,800 &&
N(\ell^+\ell^+)/N(\ell^-\ell^-) = 1.33.
\end{align}
It is advantageous to focus on cross section ratios only, so that
uncertainties for the total luminosity and the branching ratios in the decay
chains Eq.~\eqref{eq:chain} cancel out. From the measurement of the ratio
$N(\ell^+\ell^+)/N(\ell^-\ell^-)$ the Dirac theory, in comparison to the
Majorana theory, can be rejected with a statistical significance of more than
7$\,\sigma$.

However, systematic error sources are important and need to be taken into
account. Large sources for systematic uncertainties are the measurement of the
squark and gluino masses, the proton parton distribution functions (PDFs) and
missing next-to-next-to-leading order radiative corrections for the production
cross sections. Following Ref.~\cite{yuk2}, we assume $\delta m_{\gw} =
12$~GeV and $\delta m_{\tilde{q}} = 10$~GeV and derive the error from higher
order corrections from the scale dependence of the next-to-leading order
result \cite{Beenakker}.  For the PDFs we expect that the current uncertainty
will be improved by a factor of two due to the final HERA analyses for the
gluon PDF, and Tevatron and LHC data for the quark PDFs.  Including these
systematic errors, the significance is reduced to a level of about 2$\sigma$.

\begin{figure}
\vspace{-2ex}
\epsfig{figure=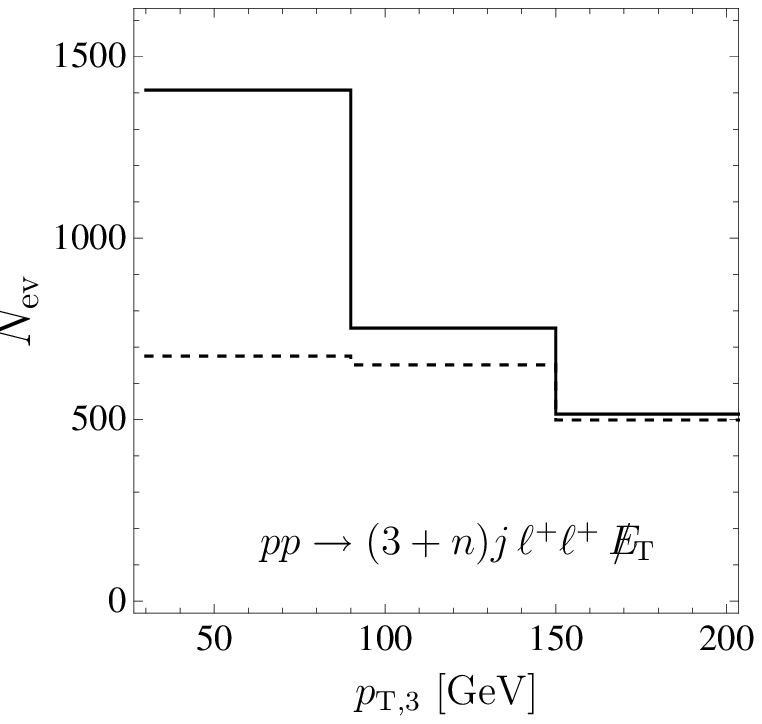, height=6cm}
\hspace{1cm}
\epsfig{figure=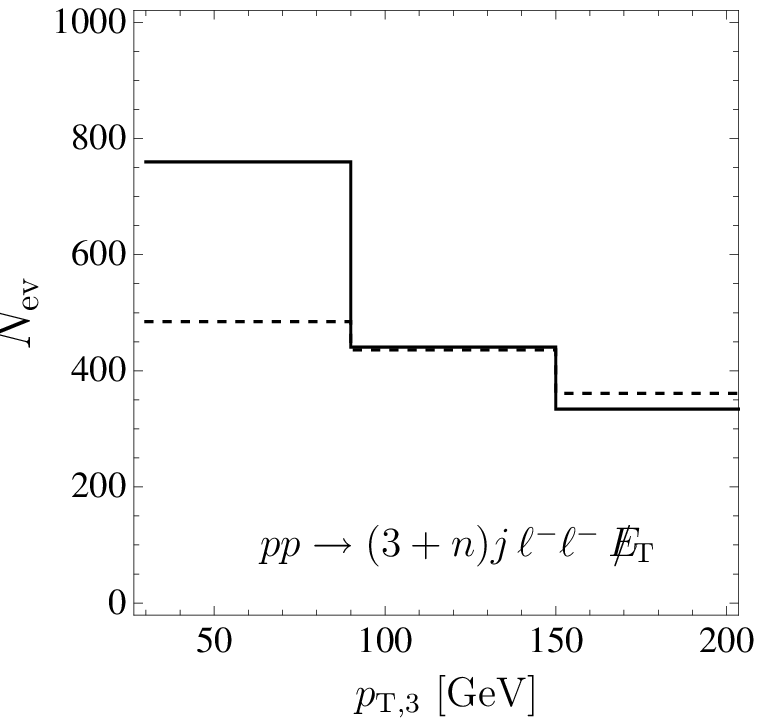, height=6cm}
\vspace{-1em}
\caption{Distribution of the transverse momentum of the third jet for the
  $\ell^+\ell^+$ (left) and the $\ell^-\ell^-$ (right) signal stemming from
  squark and/or gluino production, for Majorana theory (solid) and Dirac
  theory (dashed). The plots show the distributions in three bins, for
  SPS1a$'$ masses.} 
\label{fig:pt3}
\end{figure}

Fortunately the result can be improved considerably by considering the
distribution of the transverse momentum of the third hardest jet, $p_{\rm
  T,3}$ in the signal events, which accentuates gluino decays. Due to the
extra jet from the gluino decay $\gw \to \tilde{q}\,\bar{q}, \, \tilde{q}^*q$,
this distribution is sensitive to the relative contributions from squark pair
production, squark-gluino production, and gluino pair production. As shown in
Fig.~\ref{fig:pt3}, in the Majorana theory the $p_{\rm T,3}$-distribution is
peaked at low values of $p_{\rm T,3}$, as a result of the sizable contribution
from $\tilde{q}_L\tilde{q}_L$ production. On the other hand, the Dirac theory
predicts a relatively larger signal from the squark-gluino super-Compton
process compared to squark pair production. The gluino decay leads to a hard
third jet, so that the $p_{\rm T,3}$-distribution falls off more slowly
towards high momenta.

Dividing the $p_{\rm T,3}$-spectrum into 3 bins in the range $p_{\rm T,3}\in
[30,200]$~GeV, a fit to the distributions for the $\ell^+ \ell^+$ and $\ell^-
\ell^-$ final states allows a statistical discrimination between the Majorana
and Dirac theory with 11.3 standard deviations (for $\int \! {\mathcal{L}} = $
300 fb$^{-1}$).  Taking into account systematic errors as above, we find that
the Dirac theory can be separated from the Majorana theory by more than
10.7$\,\sigma$.

Finally, we comment on a subtle issue: If the Yukawa coupling between gluinos,
squarks and quarks is treated as an unknown parameter, as in
Ref.~\cite{yuk,yuk2}, one may worry that a non-standard value of the Yukawa
coupling mimics the effect of the Dirac theory at the LHC. We have analyzed
this problem by repeating the fit to the $p_{\rm T,3}$-spectrum with the
Yukawa coupling as a free parameter. We have also included a total cross
section measurement in this fit, with a conservative error of 30\%. It turns
out that with a free-floating Yukawa coupling the Majorana and Dirac theories
can be distinguished with a reduced statistical significance of 4.7$\sigma$
(4.5$\sigma$ including systematics).  Thus, with a medium significance, the
Yukawa coupling and the Majorana/Dirac nature of the gluinos can be determined
simultaneously and independently.

\section{SUMMARY}

\noindent
If supersymmetry is realized in nature at low energies, the next steps after
the discovery of supersymmetric particles will be the measurement of their
properties. While the measurements of masses, spins and Yukawa couplings at
the LHC has been discussed in earlier reports
\cite{weigl,Barr:2004ze,lhcspin,yuk,yuk2}, we have focused here on studies of
the Majorana nature of gluinos [and in a restricted form on neutralinos which
will be treated in depth in a later investigation].

The parallelism between self-conjugate neutral gauge bosons and their
fermionic supersymmetric partners induces the Majorana nature of these
particles in the minimal formulation of the theory.  Nevertheless,
experimental tests of the Majorana character would provide non-trivial insight
into the potential realization of supersymmetry in nature, since extended
supersymmetric models can include Dirac gauginos. N=2 supersymmetry
provides a solid theoretical basis for formulating such a testing ground.
Since the fermionic degrees of freedom are doubled in the gauge sector, the
ensuing two Majorana fields can be joined to a single Dirac field if the
masses are chosen identical. Moreover, a continuous path could be designed
connecting the original MSSM N=1 Majorana theory and the N=2 Dirac theory
by variation of mass parameters. The MSSM corresponds in this frame to a
parameter space point in which one of the N=2 mass parameters is shifted to
infinity, leading to the decoupling of the additional gaugino states. For
equal mass parameters, on the other side, the Dirac theory emerges in a
natural way.

It is interesting to note that the transition from the Majorana to the Dirac
theory  is smooth, suggesting the notion of a near-Dirac field in the approach
to the Dirac limit. This notion proves very useful in the analysis of the two
theories.

There are several methods to investigate the Majorana nature of gluinos. In
the original form, decays to heavy stop/top quarks are exploited
\cite{gluino_Majorana} to study that the final state in the fermion decay $\gw
\to \tilde{t} \bar{t} + {\tilde{t}}^\ast t$ is self-conjugate. In this report
we have explored an alternative by studying the nature of $t$-channel
exchanged gluinos. While the cross section for the scattering processes with
equal-chirality quarks $q_L q_L \to \tilde{q}_L \tilde{q}_L$ is non-zero in
the Majorana theory, it vanishes in the Dirac theory. Likewise for two
R-chiralities. However, note that two unlike-chirality quarks can generate
squarks also in the Dirac theory. L-squarks in the final state can be tagged
by measuring the lepton charges in their chargino decay modes.  Owing to the
dominance of $u$-quarks over $d$-quarks in the proton, the Majorana theory
predicts large rates of like-sign dilepton final states from squark pair
production with an excess of positively charged leptons while they are absent,
apart from a small number of remnant channels, in the Dirac theory. In a
realistic analysis one has to include gluino production processes which can
also feed the like-sign dilepton signal but can be discriminated by extra jet
emission from the gluino decays. {\it Conclusio generalis}, the Majorana
theory can be discriminated from the Dirac theory using like-sign dilepton
events at the level of more than 10$\sigma$.

In this analysis we focussed on a scenario where gluinos are somewhat heavier
than first and second generation squarks. If gluinos are much lighter, most
squarks will decay into gluinos rather than into neutralinos and charginos. In
this case one expects, {\it in toto}, approximately equal $\ell^+\ell^+$ and
$\ell^-\ell^-$ events in both the Majorana and the Dirac theory. However, the
dominant process will then be gluino pair production, which has a two times
larger cross section in the Dirac theory. In addition, left/right-chiral
correlations among top or bottom quark pairs in gluino-pair decays are
different in the Majorana and Dirac theory and they generate different
experimental signatures. While we did not perform a detailed analysis of such
a scenario, we expect that the two discriminants should allow a clean
separation of the Majorana and Dirac theories also in this case.

Similar analyses can also be designed for electroweak neutralinos. Some of
these tests can be performed at the LHC while other very clean reactions, like
$e^- e^- \to {\tilde{e}}^- {\tilde{e}}^-$, can be carried out at TeV linear
colliders. The results of these investigations will be presented in a sequel
to this report.

%
\acknowledgments{}

\noindent
We are grateful to A.~M.~Cooper-Sarkar, A.~Glazov, T.~Hebbeker, and S.~Lammel
for communications on various experimental aspects of this study.  Particular
thanks go to J.~Kalinowski for the critical reading of the manuscript. Special
thanks go to M.~M.~M\"uhlleitner and P.~Skands for clarifying issues on the
branching ratios of gluino decays.  The work by SYC was supported in part by the
Korea Research Foundation Grant funded by the Korean Government (MOERHRD, Basic
Research Promotion Fund) (KRF-2007-521-C00065) and in part by KOSEF through CHEP
at Kyungpook National University. The work of MD was partially supported by
Bundesministerium f\"ur Bildung und Forschung under contract no. 05HT6PDA, and
partially by the Marie Curie Training Research Networks ``UniverseNet'' under
contract no. MRTN-CT-2006-035863, ``ForcesUniverse'' under contract no.
MRTN-CT-2004-005104, as well as ``The Quest for Unification'' under contract no.
MRTN-CT-2004-503369.  Work at ANL is supported in part by the US DOE, Division
of HEP, Contract DE-AC-02-06CH11357. PMZ is grateful to the Inst.~Theor.~Phys.~E
for the warm hospitality extended to him at RWTH Aachen.

\vskip 0.3cm

%


\begin{thebibliography}{99}

\bibitem{Wess}
  Yu.~A.~Golfand and E.~P.~Likhtman, JETP\ Lett.\ {\bf 13} (1971) 3214;
  J.~Wess and B.~Zumino,
  Nucl.\ Phys.\  B {\bf 70} (1974) 39.

\bibitem{Nilles}
   H.~P.~Nilles,
  Phys.\ Rept.\  {\bf 110} (1984) 1;
  H.~E.~Haber and G.~L.~Kane,
  Phys.\ Rept.\  {\bf 117} (1985) 75.

\bibitem{Drees}
   M.~Drees, R.~Godbole and P.~Roy,
   ``Theory and phenomenology of sparticles: An account of four-dimensional N=1
   supersymmetry in high energy physics,''
{\it  Hackensack, USA: World Scientific (2004) 555 p};
 P.~Binetruy,
  ``Supersymmetry: Theory, experiment and cosmology,''
{\it  Oxford, UK: Oxford Univ. Pr. (2006) 520 p};
 J.~Wess and J.~Bagger,
{\it  Princeton, USA: Univ. Pr. (1992) 259 p}.

\bibitem{Benakli}
   K.~Benakli and C.~Moura, {\it in} M.~M.~Nojiri {\it et al.},
  arXiv:0802.3672 [hep-ph].

\bibitem{Fayet}
   P.~Fayet,
  Nucl.\ Phys.\  B {\bf 113} (1976) 135;
L.~\'Alvarez-Gaum\'e and S.~F.~Hassan,
  Fortsch.\ Phys.\  {\bf 45} (1997) 159
  [arXiv:hep-th/9701069].

\bibitem{Majorana_to_Dirac} See, e.g.,
  S.~M.~Bilenky and S.~T.~Petcov,
  Rev.\ Mod.\ Phys.\  {\bf 59} (1987) 671
  [Erratum-ibid.\  {\bf 61} (1989) 169];
R.~N.~Mohapatra and P.~B.~Pal,
  World Sci.\ Lect.\ Notes Phys.\  {\bf 72} (2004) 1.

\bibitem{LHC}
  A.~Airapetian {\it et al.},
  {\it ATLAS Detector and physics performance technical design report, Vol.\ 1}, 
  ATLAS-TDR-14, CERN/LHCC-99-14;
  G.L. Bayatian {\it et al.}, 
  {\it CMS technical design report, volume II: Physics performance},
  J.\ Phys.\ G {\bf 34}, 995 (2007).

\bibitem{LC}
  E.~Accomando {\it et al.},
  Phys.\ Rept.\  {\bf 299} (1998) 1
  [arXiv:hep-ph/9705442];
  J.~A.~Aguilar-Saavedra {\it et al.}  [ECFA/DESY LC Physics Working Group],
  {\it TESLA Technical Design Report Part III: Physics at an e+e- Linear
  Collider},
  arXiv:hep-ph/0106315;
  J.~Brau {\it et al.}  [ILC Collaboration],
  {\it ILC Reference Design Report, Vol. 1 -- Executive Summary},
  arXiv:0712.1950 [physics.acc-ph];
  A.~Djouadi {\it et al.},
  {\it ILC Reference Design Report, Vol.2},
  arXiv:0709.1893 [hep-ph];
  E.~Accomando {\it et al.}  [CLIC Physics Working Group],
  {\it Physics at the CLIC multi-TeV linear collider},
  arXiv:hep-ph/0412251.

\bibitem{yuk}
  A.~Freitas and P.~Z.~Skands,
  JHEP {\bf 0609} (2006) 043;
  {\it see also} A.~Brandenburg, M.~Maniatis, M.~M.~Weber and P.~M.~Zerwas,
  arXiv:0806.3875 [hep-ph].

\bibitem{yuk2}
  A.~Freitas, P.~Z.~Skands, M.~Spira and P.~M.~Zerwas, 
  JHEP {\bf 0707} (2007) 025, [arXiv:hep-ph/0703160].

\bibitem{no}
 M.~M.~Nojiri and M.~Takeuchi,
  Phys.\ Rev.\  D {\bf 76} (2007) 015009
  [arXiv:hep-ph/0701190].

\bibitem{gluino_Majorana}
  R.~M.~Barnett, J.~F.~Gunion and H.~E.~Haber,
  Phys.\ Lett.\  B {\bf 315} (1993) 349
  [arXiv:hep-ph/9306204];
S.~Kraml and A.~R.~Raklev,
  Phys.\ Rev.\  D {\bf 73} (2006) 075002
  [arXiv:hep-ph/0512284];
  A.~Alves, O.~Eboli and T.~Plehn,
  Phys.\ Rev.\  D {\bf 74} (2006) 095010
  [arXiv:hep-ph/0605067].

\bibitem{Tev}
 A.~Abulencia et al., The CDF
 Collaboration, Phys.\ Rev.\ Lett.\ {\bf 96} (2006) 171802; and
 \url{http://www-cdf.fnal.gov/physics/exotic/r2a/20080410.bbmet\_gluinosbottom/}.

\bibitem{alwall}
J.~Alwall, D.~Rainwater and T.~Plehn,
  Phys.\ Rev.\  D {\bf 76} (2007) 055006
  [arXiv:0706.0536 [hep-ph]].

\bibitem{neutralino_decay_Majorana}
  S.~Y.~Choi and Y.~G.~Kim,
  Phys.\ Rev.\  D {\bf 69} (2004) 015011
  [arXiv:hep-ph/0311037];
  S.~Y.~Choi, B.~C.~Chung, J.~Kalinowski, Y.~G.~Kim and K.~Rolbiecki,
  Eur.\ Phys.\ J.\  C {\bf 46} (2006) 511
  [arXiv:hep-ph/0504122].

\bibitem{supersoft}
  P.~J.~Fox, A.~E.~Nelson and N.~Weiner,
  JHEP {\bf 0208}, 035 (2002)
  [arXiv:hep-ph/0206096].

\bibitem{QA}
  I.~Antoniadis, K.~Benakli, A.~Delgado and M.~Quiros,
  CERN-PH-TH-2006-188
  [arXiv:hep-ph/0610265].

\bibitem{Chacko:2004mi}
  Z.~Chacko, P.~J.~Fox and H.~Murayama,
  Nucl.\ Phys.\  B {\bf 706} (2005) 53
  [arXiv:hep-ph/0406142].

\bibitem{dterm}
  Y.~Nomura, D.~Poland and B.~Tweedie,
  Nucl.\ Phys.\  B {\bf 745}, 29 (2006)
  [arXiv:hep-ph/0509243].

\bibitem{rsymm}
  L.~J.~Hall and L.~Randall,
  Nucl.\ Phys.\  B {\bf 352}, 289 (1991);
  G.~D.~Kribs, E.~Poppitz and N.~Weiner,
  arXiv:0712.2039 [hep-ph].

\bibitem{CHKZ}
  S.~Y.~Choi, H.~E.~Haber, J.~Kalinowski and P.~M.~Zerwas,
  Nucl.\ Phys.\  B {\bf 778} (2007) 85
  [arXiv:hep-ph/0612218].

\bibitem{Beenakker}
  W.~Beenakker, R.~H\"opker, M.~Spira and P.~M.~Zerwas,
  Nucl.\ Phys.\  B {\bf 492} (1997) 51
  [arXiv:hep-ph/9610490].

\bibitem{LlSmith}
P.~R.~Harrison and C.~H.~Llewellyn Smith,
  Nucl.\ Phys.\  B {\bf 213} (1983) 223
  [Erratum-ibid.\  B {\bf 223} (1983) 542];
S.~Dawson, E.~Eichten and C.~Quigg,
  Phys.\ Rev.\  D {\bf 31} (1985) 1581.

\bibitem{Kul}
A.~Kulesza and L.~Motyka,
  arXiv:0807.2405 [hep-ph].

\bibitem{bddk}
  S.~Bornhauser, M.~Drees, H.~K.~Dreiner and J.~S.~Kim,
  Phys.\ Rev.\  D {\bf 76}, 095020 (2007)
  [arXiv:0709.2544 [hep-ph]].

\bibitem{hollik}
W. Hollik and E. Mirabella, arXiv:0806.1433 [hep-ph].

\bibitem{Beenakker2}
  W.~Beenakker, R.~H\"opker and P.~M.~Zerwas,
  Phys.\ Lett.\  B {\bf 378} (1996) 159
  [arXiv:hep-ph/9602378].

\bibitem{Djouadi}
  M.~M\"uhlleitner, A.~Djouadi and Y.~Mambrini,
  Comput.\ Phys.\ Commun.\  {\bf 168} (2005) 46
  [arXiv:hep-ph/0311167].

\bibitem{Barr:2004ze}
  A.~J.~Barr,
  Phys.\ Lett.\  B {\bf 596} (2004) 205
  [arXiv:hep-ph/0405052].

\bibitem{lhcspin}
  J.~M.~Smillie and B.~R.~Webber,
  JHEP {\bf 0510} (2005) 069
  [arXiv:hep-ph/0507170];
  D.~J.~Miller, P.~Osland and A.~R.~Raklev,
  JHEP {\bf 0603} (2006) 034
  [arXiv:hep-ph/0510356].

\bibitem{SPS1a}
  J.~A.~Aguilar-Saavedra {\it et al.},
  Eur.\ Phys.\ J.\  C {\bf 46} (2006) 43
  [arXiv:hep-ph/0511344].

\bibitem{Keung:1983nq}
  W.~Y.~Keung and L.~Littenberg,
  Phys.\ Rev.\  D {\bf 28}, 1067 (1983).

\bibitem{gudi}
  G.~A.~Moortgat-Pick {\it et al.},
  Phys.\ Rept.\  {\bf 460} (2008) 131
  [arXiv:hep-ph/0507011].

\bibitem{slep}
  A.~Freitas, A.~von Manteuffel and P.~M.~Zerwas,
  Eur.\ Phys.\ J.\  C {\bf 34} (2004) 487
  [arXiv:hep-ph/0310182].

\bibitem{AguilarSaavedra:2003hw}
  J.~A.~Aguilar-Saavedra and A.~M.~Teixeira,
  Nucl.\ Phys.\  B {\bf 675} (2003) 70
  [arXiv:hep-ph/0307001].

\bibitem{SPS}
B.~C.~Allanach {\it et al.},
``The Snowmass points and slopes: Benchmarks for SUSY searches,''
{\it in} Proc. of the APS/DPF/DPB Summer Study on the Future of Particle
Physics, Snowmass (Colorado) 2001, 
and Eur.\ Phys.\ J.\  C {\bf 25} (2002) 113 [arXiv:hep-ph/0202233].

\bibitem{ellisea}
   J.~R.~Ellis, S.~Heinemeyer, K.~A.~Olive, A.~M.~Weber and G.~Weiglein,
  JHEP {\bf 0708} (2007) 083
  [arXiv:0706.0652 [hep-ph]].

\bibitem{pythia}
  T.~Sj\"ostrand, S.~Mrenna and P.~Skands,
  JHEP {\bf 0605}, 026 (2006)
  [arXiv:hep-ph/0603175].

\bibitem{weigl}
  I.~Hinchliffe, F.~E.~Paige, M.~D.~Shapiro, J.~S\"oderqvist and W.~Yao,
  Phys.\ Rev.\  D {\bf 55}, 5520 (1997)
  [arXiv:hep-ph/9610544];
  C.~G.~Lester and D.~J.~Summers,
  Phys.\ Lett.\  B {\bf 463}, 99 (1999)
  [arXiv:hep-ph/9906349];
    G.~Weiglein {\it et al.}  [LHC/LC Study Group],
  Phys.\ Rept.\  {\bf 426} (2006) 47
  [arXiv:hep-ph/0410364].

\end{thebibliography}
\end{document}